\documentclass[ALICE,manyauthors]{cernphprep}
\usepackage[comma,square,numbers,sort&compress]{natbib}
\usepackage[T1]{fontenc}
\usepackage{hyperref}
\usepackage{lineno}
\usepackage{xcolor}
\usepackage{xspace}
\usepackage{multirow}
\usepackage{makecell}
\usepackage{booktabs}
\usepackage{upgreek}
\usepackage{comment}
\usepackage{amsmath}
\usepackage{orcidlink}
\begin{document}
%

\newcommand{\ee}               {\ensuremath{\mathrm{e^+e^-}}\xspace} 
\newcommand{\pp}               {\ensuremath{\mathrm{pp}}\xspace}
\newcommand{\ppbar}            {\ensuremath{\mathrm{p\overline{p}}}\xspace}
\newcommand{\pPb}              {\ensuremath{\mathrm{p--Pb}}\xspace}
\newcommand{\PbPb}             {\ensuremath{\mathrm{Pb--Pb}}\xspace}

\newcommand{\av}[1]            {\ensuremath{\left\langle #1 \right\rangle}\xspace}

\newcommand{\s}                {\ensuremath{\sqrt{s}}\xspace}
\newcommand{\RAA}              {\ensuremath{R_\mathrm{AA}}\xspace}
\newcommand{\TAA}              {\ensuremath{T_\mathrm{AA}}\xspace}
\newcommand{\pt}               {\ensuremath{p_\mathrm{T}}\xspace}
\newcommand{\mt}               {\ensuremath{m_\mathrm{T}}\xspace}
\newcommand{\de}               {\ensuremath{\mathrm{d}}\xspace}
\newcommand{\dEdx}             {\ensuremath{\de E/\de x}\xspace}
\newcommand{\dNdpt}            {\ensuremath{\de N/\de\pt}\xspace}
\newcommand{\dNdptdy}          {\ensuremath{\de^{2} N/\de\pt\de y}\xspace}
\newcommand{\dNdy}             {\ensuremath{\de N/\de y}\xspace}
\newcommand{\mur}              {\ensuremath{\mu_\mathrm{R}}\xspace}
\newcommand{\muf}              {\ensuremath{\mu_\mathrm{F}}\xspace}
\newcommand{\fprompt}          {\ensuremath{f_\mathrm{prompt}}\xspace}
\newcommand{\fnonprompt}       {\ensuremath{f_\mathrm{non\text{-}prompt}}\xspace}
\newcommand{\fnonpromptraw}    {\ensuremath{f^\mathrm{raw}_\mathrm{non\text{-}prompt}}\xspace}
\newcommand{\fnonpromptLow}       {\ensuremath{f^{30-100\%}_\mathrm{non\text{-}prompt}}\xspace}
\newcommand{\fnonpromptMid}       {\ensuremath{f^{0.1-30\%}_\mathrm{non\text{-}prompt}}\xspace}
\newcommand{\fnonpromptHig}       {\ensuremath{f^{0-0.1\%}_\mathrm{non\text{-}prompt}}\xspace}
\newcommand{\fnptofnpMB}      {\ensuremath{f^\mathrm{mult}_\mathrm{non\text{-}prompt}/f^{\rm{INEL>0}}_\mathrm{non\text{-}prompt}}\xspace}
\newcommand{\fnpLowtofnpMB}      {\ensuremath{f^{30-100\%}_\mathrm{non\text{-}prompt}/f^{\rm{INEL>0}}_\mathrm{non\text{-}prompt}}\xspace}
\newcommand{\fnpMidtofnpMB}      {\ensuremath{f^{0.1-30\%}_\mathrm{non\text{-}prompt}/f^{\rm{INEL>0}}_\mathrm{non\text{-}prompt}}\xspace}
\newcommand{\fnpHigtofnpMB}      {\ensuremath{f^{0-0.1\%}_\mathrm{non\text{-}prompt}/f^{\rm{INEL>0}}_\mathrm{non\text{-}prompt}}\xspace}

\newcommand{\f}[1]             {\ensuremath{f_\mathrm{#1}}\xspace}
\newcommand{\rawY}[1]          {\ensuremath{Y_\mathrm{#1}}\xspace}
\newcommand{\AccEff}           {\ensuremath{(\mathrm{Acc} \times \epsilon)}\xspace}
\newcommand{\AccEffNP}         {\ensuremath{(\mathrm{Acc}\times\epsilon)_{\mathrm{non\text{-}prompt}}}\xspace}
\newcommand{\AccEffP}          {\ensuremath{(\mathrm{Acc}\times\epsilon)_{\mathrm{prompt}}}\xspace}
\newcommand{\Np}               {\ensuremath{N_\mathrm{prompt}}\xspace}
\newcommand{\Nnp}              {\ensuremath{N_\mathrm{non\text{-}prompt}}\xspace}

\newcommand{\MeV}              {\ensuremath{\mathrm{MeV}}\xspace}
\newcommand{\GeV}              {\ensuremath{\mathrm{GeV}}\xspace}
\newcommand{\TeV}              {\ensuremath{\mathrm{TeV}}\xspace}
\newcommand{\fm}               {\ensuremath{\mathrm{fm}}\xspace}
\newcommand{\mm}               {\ensuremath{\mathrm{mm}}\xspace}
\newcommand{\cm}               {\ensuremath{\mathrm{cm}}\xspace}
\newcommand{\m}                {\ensuremath{\mathrm{m}}\xspace}
\newcommand{\mum}              {\ensuremath{\mathrm{\upmu m}}\xspace}
\newcommand{\ns}               {\ensuremath{\mathrm{ns}}\xspace}
\newcommand{\mrad}             {\ensuremath{\mathrm{mrad}}\xspace}
\newcommand{\mb}               {\ensuremath{\mathrm{mb}}\xspace}
\newcommand{\mub}              {\ensuremath{\mathrm{\upmu b}}\xspace}
\newcommand{\lumi}             {\ensuremath{\mathcal{L}_\mathrm{int}}\xspace}
\newcommand{\nbinv}            {\ensuremath{\mathrm{nb^{-1}}}\xspace}
\newcommand{\mubinv}           {\ensuremath{\mathrm{\upmu b^{-1}}}\xspace}

\newcommand{\pip}              {\ensuremath{\mathrm{\uppi^{+}}}\xspace}
\newcommand{\pim}              {\ensuremath{\mathrm{\uppi^{-}}}\xspace}
\newcommand{\kap}              {\ensuremath{\mathrm{\rm{K}^{+}}}\xspace}
\newcommand{\kam}              {\ensuremath{\mathrm{\rm{K}^{-}}}\xspace}
\newcommand{\pbar}             {\ensuremath{\mathrm{\rm\overline{p}}}\xspace}
\newcommand{\kzero}            {\ensuremath{\mathrm{K^0_S}}\xspace}
\newcommand{\Lam}              {\ensuremath{\mathrm{\Lambda}}\xspace}
\newcommand{\antiLam}          {\ensuremath{\mathrm{\overline{\Lambda}}}\xspace}
\newcommand{\Om}               {\ensuremath{\mathrm{\Omega^-}}\xspace}
\newcommand{\antiOm}           {\ensuremath{\mathrm{\overline{\Omega}^+}}\xspace}
\newcommand{\X}                {\ensuremath{\mathrm{\Xi^-}}\xspace}
\newcommand{\antiX}            {\ensuremath{\mathrm{\overline{\Xi}^+}}\xspace}
\newcommand{\Xipm}             {\ensuremath{\mathrm{\Xi^{\pm}}}\xspace}
\newcommand{\Opm}              {\ensuremath{\mathrm{\Omega^{\pm}}}\xspace}
\newcommand{\DzerotoKpi}       {\ensuremath{\mathrm{D^0 \to K^-\uppi^+}}\xspace}
\newcommand{\DplustoKpipi}     {\ensuremath{\mathrm{D^+\to K^-\uppi^+\uppi^+}}\xspace}
\newcommand{\DstartoDpi}       {\ensuremath{\mathrm{D^{*+} \to \rm D^0 \uppi^+}}\xspace}
\newcommand{\Dstophipi}        {\ensuremath{\mathrm{D_s^+\to \upphi\uppi^+}}\xspace}
\newcommand{\Dstophipipm}      {\ensuremath{\mathrm{D_s^\pm\to \upphi\uppi^\pm}}\xspace}
\newcommand{\DstophipitoKKpi}  {\ensuremath{\mathrm{D_s^+\to \upphi\uppi^+\to K^-K^+\uppi^+}}\xspace}
\newcommand{\DplustoKKpi}      {\ensuremath{\mathrm{D^+\to K^-K^+\uppi^+}}\xspace}
\newcommand{\phitoKK}          {\ensuremath{\mathrm{\upphi\to  K^-K^+}}\xspace}
\newcommand{\DstoKzerostarK}   {\ensuremath{\mathrm{D_s^+\to \overline{K}^{*0} K^+}}\xspace}
\newcommand{\Dstofzeropi}      {\ensuremath{\mathrm{D_s^+\to f_0(980) \uppi^+}}\xspace}
\newcommand{\fzero}            {\ensuremath{\mathrm{f_0(980)}}\xspace}
\newcommand{\Kzerostar}        {\ensuremath{\mathrm{\overline{K}^{*0}}}\xspace}
\newcommand{\Dzero}            {\ensuremath{\mathrm{D^0}}\xspace}
\newcommand{\Dzerobar}         {\ensuremath{\mathrm{\overline{D}\,^0}}\xspace}
\newcommand{\Dstar}            {\ensuremath{\mathrm{D^{*+}}}\xspace}
\newcommand{\Dstarm}           {\ensuremath{\mathrm{D^{*-}}}\xspace}
\newcommand{\DstarZero}        {\ensuremath{\mathrm{D^{*0}}}\xspace}
\newcommand{\DstarS}           {\ensuremath{\mathrm{D_s^{*+}}}\xspace}
\newcommand{\Dplus}            {\ensuremath{\mathrm{D^+}}\xspace}
\newcommand{\Dminus}           {\ensuremath{\mathrm{D^-}}\xspace}
\newcommand{\Ds}               {\ensuremath{\mathrm{D_s^+}}\xspace}
\newcommand{\Dspm}             {\ensuremath{\mathrm{D_s^\pm}}\xspace}
\newcommand{\Dsstar}           {\ensuremath{\mathrm{D_s^{*+}}}\xspace}
\newcommand{\LambdaC}          {\ensuremath{\Lambda_\mathrm{c}^+}\xspace}
\newcommand{\KKpi}             {\ensuremath{\mathrm{K^-K^+\uppi^+}}\xspace}
\newcommand{\cubar}            {\ensuremath{\mathrm{c\bar{u}}}\xspace}
\newcommand{\cdbar}            {\ensuremath{\mathrm{c\bar{d}}}\xspace}
\newcommand{\ccbar}            {\ensuremath{\mathrm{c\overline{c}}}\xspace}
\newcommand{\bbbar}            {\ensuremath{\mathrm{b\overline{b}}}\xspace}
\newcommand{\Bzero}            {\ensuremath{\mathrm{B^0}}\xspace}
\newcommand{\Bplus}            {\ensuremath{\mathrm{B^+}}\xspace}
\newcommand{\Bzeroplus}        {\ensuremath{\mathrm{B^{0,+}}}\xspace}
\newcommand{\Bs}               {\ensuremath{\mathrm{B_s^0}}\xspace}
\newcommand{\Lambdab}          {\ensuremath{\mathrm{\Lambda_b^0}}\xspace}
\newcommand{\Jpsi}             {\ensuremath{\mathrm{J}/\uppsi}\xspace}
\newcommand{\psitwo}           {\ensuremath{\uppsi(2\mathrm{S})}\xspace}
\newcommand{\Chic}             {\ensuremath{\upchi_\mathrm{c1}(3872)}\xspace}
\newcommand{\Vdecay} 	       {\ensuremath{\mathrm{V^{0}}}\xspace}
\newcommand{\bhad}             {\ensuremath{\mathrm{H_b}}\xspace}
\newcommand{\Ztobbbar}         {\ensuremath{\mathrm{Z\to b\overline{b}}}\xspace}
\newcommand{\fctoD}            {\ensuremath{f(\mathrm{c}\to\mathrm{D})}\xspace}
\newcommand{\fbtoB}            {\ensuremath{f(\mathrm{b}\to\mathrm{B})}\xspace}
\newcommand{\fctoHc}           {\ensuremath{f(\mathrm{c}\to\mathrm{H_c})}\xspace}
\newcommand{\fbtoHb}           {\ensuremath{f(\mathrm{b}\to\mathrm{H_b})}\xspace}
\newcommand{\fbtoHc}           {\ensuremath{f(\mathrm{b}\to\mathrm{H_c})}\xspace}
\newcommand{\BtoD}             {\ensuremath{\mathrm{b}\to\mathrm{D+X}}\xspace}
\newcommand{\dndeta}           {\ensuremath{\mathrm{d}N_\mathrm{ch}/\mathrm{d}\eta}\xspace}
\newcommand{\avdndeta}         {\ensuremath{\langle\dndeta\rangle}\xspace}

\newcommand{\effNP}[1]         {\ensuremath{(\mathrm{Acc}\times\epsilon)^\mathrm{non\text{-}prompt}_\mathrm{#1}}\xspace}
\newcommand{\effP}[1]          {\ensuremath{(\mathrm{Acc}\times\epsilon)^\mathrm{prompt}_\mathrm{#1}}\xspace}
\newcommand{\eff}              {\ensuremath{\mathrm{Acc}\times\epsilon}\xspace}

\newcommand{\inelgtrz}         {\ensuremath{\text{INEL} > 0}\xspace}
\newcommand{\Lint}             {\ensuremath{\mathcal{L}_\mathrm{int}}\xspace}

\newcommand{\fonll}            {\textsc{FONLL}\xspace}
\newcommand{\gmvfns}           {\textsc{GM-VFNS}\xspace}
\newcommand{\pythia}           {\textsc{PYTHIA~8}\xspace}
\newcommand{\pythiaprec}       {\textsc{PYTHIA~8.243}\xspace}
\newcommand{\epos}             {\textsc{EPOS}\xspace}
\newcommand{\eposthree}        {\textsc{EPOS~3}\xspace}
\newcommand{\eposfour}         {\textsc{EPOS~4}\xspace}
\newcommand{\cgc}              {\textsc{CGC}\xspace}
\newcommand{\herwig}           {\textsc{HERWIG~7}\xspace}
\newcommand{\clr}              {\textsc{CLR-BLC}\xspace}
\newcommand{\geant}            {\textsc{GEANT3}\xspace}

\begin{titlepage}
\PHyear{2023}       
\PHnumber{018}      
\PHdate{14 February}  

\title{Measurement of the non-prompt D-meson fraction as a function of multiplicity in proton--proton collisions at $\pmb{\s = 13~\TeV}$}
\ShortTitle{Non-prompt D-meson fraction as a function of multiplicity in pp at $\s = 13~\TeV$}   

\Collaboration{ALICE Collaboration\thanks{See Appendix~\ref{app:collab} for the list of collaboration members}}
\ShortAuthor{ALICE Collaboration} 

\begin{abstract}
The fractions of non-prompt (i.e.~originating from beauty-hadron decays) \Dzero and \Dplus mesons with respect to the inclusive yield are measured as a function of the charged-particle multiplicity in proton--proton collisions at a centre-of-mass energy of $\s=13~\TeV$ with the ALICE detector at the LHC. The results are reported in intervals of transverse momentum (\pt) and integrated in the range $1<\pt<24~\GeV/c$. The fraction of non-prompt \Dzero and \Dplus mesons is found to increase slightly as a function of \pt in all the measured multiplicity intervals, while no significant dependence on the charged-particle multiplicity is observed. In order to investigate the production and hadronisation mechanisms of charm and beauty quarks, the results are compared to \pythia as well as \epos 3 and \epos 4  Monte Carlo simulations, and to calculations based on the colour glass condensate including three-pomeron fusion.

\end{abstract}
\end{titlepage}

\setcounter{page}{2} 

\section{Introduction}
\label{sec:intro}

Measurements of the production of hadrons containing heavy quarks, i.e. charm or beauty, in proton--proton (pp) collisions provide an important test of quantum chromodynamics (QCD) calculations. Several measurements of charm- and beauty-hadron production were carried out in pp collisions by the ALICE
~\cite{ALICE:2019nxm,ALICE:2021mgk,ALICE:2021edd,ALICE:2019nuy,ALICE:2019rmo,ALICE:2021psx,ALICE:2021bli,ALICE:2020wfu,ALICE:2020wla,ALICE:2021rzj,ALICE:2022cop}, 
ATLAS~\cite{ATLAS:2012sfc,ATLAS:2015igt,ATLAS:2013cia,ATLAS:2015esn,ATLAS:2019jpi}, CMS~\cite{CMS:2012xsp,CMS:2017qjw,CMS:2017uoy,CMS:2019uws,CMS:2016plw,CMS:2018bwt,CMS:2014oqy}, and LHCb~\cite{LHCb:2013vjr,LHCb:2016ikn,LHCb:2015swx,LHCb:2014mvo,LHCb:2015qvk,LHCb:2016qpe,LHCb:2017vec,LHCb:2019sxa,LHCb:2019fns,LHCb:2019qed} experiments at the LHC, and by the STAR experiment at RHIC~\cite{STAR:2012nbd}. The measured D- and B-meson production cross sections are generally compatible within uncertainties with theoretical predictions based on the factorisation approach, which describe them as the convolution of the parton distribution functions (PDFs), the partonic cross section calculated with perturbative QCD (pQCD) calculations, and the fragmentation functions (FFs). Calculations of the partonic cross sections are nowadays available at next-to-leading-order accuracy (like $k_\mathrm{T}$-factorisation~\cite{Maciula:2013wg,Maciula:2018iuh,Guiot:2018kfy}) or next-to-leading-order with next-to-leading logarithm resummation (like \fonll~\cite{Cacciari:1998it,Cacciari:2001td,Cacciari:2012ny} and \gmvfns~\cite{Kniehl:2004fy,Kniehl:2012ti,Benzke:2017yjn,Kramer:2017gct,Helenius:2018uul,Bolzoni:2013vya}). The FFs are typically constrained from measurements carried out in \ee or ep collisions~\cite{Braaten:1994bz}, under the assumption that the hadronisation of heavy quarks into hadrons is a universal process independent of the colliding system. However, measurements of baryons containing heavy quarks at hadronic colliders showed an enhancement of the baryon-to-meson yield ratios relative to the values measured at \ee colliders~\cite{ALICE:2021dhb,LHCb:2019fns}, challenging the assumption of the universality of the fragmentation across different collision systems. Monte Carlo (MC) generators that implement the transition from the heavy quark to the hadron via string fragmentation (as \pythia~\cite{Sjostrand:2014zea} with the  Monash-13~\cite{Skands:2014pea} tune) or cluster hadronisation (such as \herwig~\cite{Bellm:2015jjp}), in which the heavy-quark fragmentation is tuned to \ee and ep measurements, cannot reproduce the baryon-to-meson yield ratios measured in pp collisions. When including the colour reconnection mechanism beyond the leading colour (CR-BLC) approximation in \pythia~\cite{Christiansen:2015yqa}, which introduces new colour-reconnection topologies that fragment into baryons, a much better agreement with data is obtained~\cite{ALICE:2021rzj,ALICE:2020wfu,ALICE:2020wla}. In particular, three settings (`Modes' 0, 2, and 3), characterised by different constraints on the time dilation and causality, were defined in Ref.~\cite{Christiansen:2015yqa}. The time parameters are relevant in this model, because two string pieces must be able to resolve each other during the time between formation and hadronisation to reconnect, taking time-dilation effects caused by relative boosts into account. However, in case of charm baryons with strange-quark content, a significant discrepancy still remains with data even when considering the CR-BLC modes, suggesting that additional effects should be introduced in order to have a complete description of the hadronisation processes~\cite{ALICE:2021bli,ALICE:2021psx,ALICE:2022cop}. In the light-flavour sector, it was observed that increasing the string tension (`colour ropes' tune), which leads to an increase of strangeness production, a better agreement with data for the charged-particle multiplicity dependence of multi-strange hadron production is obtained~\cite{ALICE:2019avo,Bierlich:2014xba}.

Given that the production of heavy quarks occurs in initial hard partonic scattering processes while the production of light particles in the underlying event is dominated by soft processes, the measurement of heavy-flavour hadron production as a function of the charged-particle multiplicity has the potential to give insights into the interplay between the soft and hard mechanisms in particle production. In particular, multi-parton interactions (MPI)~\cite{Sjostrand:1987su,Bartalini:2009quk}, i.e.~several hard partonic interactions occurring in a single pp collision, influence the production of light quarks and gluons, affecting the total event multiplicity, as well as the production of heavy quarks. In addition, high-multiplicity events allow one to test the heavy-flavour hadron production at small Bjorken-$x$, i.e.~a kinematic region where the density of low-momentum gluons in the colliding protons is very high and is expected to reach saturation, which otherwise would require significantly larger energies~\cite{Schmidt:2020fgn}. A faster-than-linear increase has been observed in the production of prompt D mesons, as well as that of inclusive, prompt, and non-prompt (from beauty-hadron decays) \Jpsi mesons at midrapidity as a function of the charged-particle multiplicity in pp collisions~\cite{ALICE:2015ikl,ALICE:2020msa}. The same behaviour was obtained using a multiplicity estimators based on particles measured in the same pseudorapidity interval and introducing a pseudorapidity gap with respect to the heavy-flavour hadron~\cite{ALICE:2020msa}. A linear increase was instead observed in the measurement of \Jpsi mesons a forward rapidity, if a pseudorapidity gap is introduced between the \Jpsi mesons and the multiplicity estimator~\cite{ALICE:2021zkd}. This behaviour is described by several MC generators including MPI, such as \pythia~\cite{Sjostrand:2014zea} and \epos 3~\cite{Werner:2013tya}. \epos is an event generator suited for various hadronic colliding systems, from pp to nucleus--nucleus. This event generator assumes initial conditions generated in the Gribov-Regge multiple scattering framework, possibly followed by a hydrodynamical evolution applicable to all collision systems. Initial conditions are generated in the Gribov-Regge multiple scattering framework. Individual scatterings are referred to as Pomerons, and are identified with parton ladders. Each parton ladder is composed of a pQCD hard process with initial- and final-state radiation. Non-linear effects are considered by means of a saturation scale. The hadronisation is performed with a string fragmentation procedure, consisting in the decay of plasma droplets which conserves energy, momentum, and flavour. Other models based on a colour glass condensate (\cgc) with the three-pomeron fusion mechanism~\cite{Schmidt:2020fgn} are also able to describe the multiplicity dependence of the production yield of heavy-flavour hadrons~\cite{ALICE:2020msa,Schmidt:2020fgn}.

It is also important to note that the charged-particle densities reached in high-multiplicity pp collisions at LHC energies are comparable with those measured in peripheral heavy-ion collisions. Measurements in high-multiplicity pp collisions showed features that resemble those associated with the formation of a colour-deconfined state of the matter called quark--gluon plasma~\cite{Braun-Munzinger:2015hba} in heavy-ion collisions~\cite{CMS:2010ifv,ATLAS:2015hzw,ALICE:2016fzo}. In this context, one of the most interesting effects is the modification of the hadronisation mechanism. Model calculations based on statistical hadronisation~\cite{Chen:2020drg} or hadronisation via coalescence~\cite{Minissale:2020bif,Plumari:2017ntm} predict an enhancement of the baryon-to-meson and strange-to-nonstrange yield ratios as a function of the charged-particle multiplicity. The first category of models is based on the evaluation of the population of hadron states according to statistical weights governed by the masses of the hadrons and a universal temperature, while the second ones implement the recombination of partons close in phase space into the final hadrons. Recently, the ALICE Collaboration observed a multiplicity dependence of the transverse momentum (\pt) differential $\LambdaC/\Dzero$ ratio, smoothly evolving from pp to Pb--Pb collisions. The same quantity measured \pt integrated was found not to vary significantly as a function of the charged-particle multiplicity. No modification of the $\Ds/\Dzero$ ratio with increasing multiplicity was measured in pp collisions~\cite{ALICE:2021npz,ALICE:2021bib}.
Conversely, in the beauty sector, the LHCb Collaboration found evidence of an increase of the $\Bs/\Bzero$ production ratio with the multiplicity, in case of charged-particle multiplicity estimated with tracks in the same pseudorapidity interval of the B mesons~\cite{LHCb:2022syj}, while no measurements of beauty-baryon production as a function of charged-particle multiplicity are available. Finally, the fraction of \Chic and \psitwo states promptly produced at the collision vertex was found by the LHCb Collaboration to decrease as charged-particle multiplicity increases~\cite{LHCb:2020sey}. This suppression is interpreted as a consequence of the heavy-quark breakup via interactions with comoving hadrons~\cite{Esposito:2020ywk,Braaten:2020iqw}.

In this article, the first measurement of the fraction of \Dzero and \Dplus mesons originating from beauty-hadron decays (\fnonprompt) at midrapidity ($|y|<0.5$) is reported as a function of the charged-particle multiplicity in pp collisions at $\s=13~\TeV$. In addition, the ratio between the fraction measured in different multiplicity classes divided by the one measured in the multiplicity-integrated sample is presented. The experimental apparatus and the multiplicity determination are described in Section~\ref{sec:detector}. The measurement of \fnonprompt in six transverse momentum intervals and integrated in $1<\pt<24~\GeV/c$ is described in Section~\ref{sec:analysis}, while the evaluation of the systematic uncertainties is discussed in Section~\ref{sec:syst}. Finally, the results are presented and compared to model calculations in Section~\ref{sec:results}.
\section{Experimental apparatus and data sample}
\label{sec:detector}
The ALICE apparatus is composed of several detectors for particle reconstruction and identification at midrapidity, embedded in a large solenoidal magnet that provides a magnetic field of $B = 0.5~\mathrm{T}$ parallel to the beams. It also includes a forward muon spectrometer ($-4<\eta<-2.5$) and a set of forward and backward detectors for triggering and event characterisation. A comprehensive description of the ALICE detector and its performance is reported in Refs.~\cite{ALICE:2014sbx,ALICE:2008ngc}. 

The Inner Tracking System (ITS), consisting of six cylindrical layers of silicon detectors, allows for a precise reconstruction of primary and secondary vertices, and it is used for tracking. The Time Projection Chamber (TPC) provides up to 159 space points to reconstruct the charged-particle trajectory, and provides particle identification (PID) via the measurement of the specific ionisation energy loss \dEdx of charged particles. The Time-Of-Flight detector (TOF) extends the PID capability by measuring the flight time of charged particles from the interaction point to the TOF. These detectors cover the full azimuth in the pseudorapidity interval $|\eta|<0.9$.  The V0 detector arrays, covering the intervals $2.8 < \eta < 5.1$ (V0A) and $-3.7 < \eta < -1.7$ (V0C), are used for triggering purposes and event multiplicity measurements.

The data used for this analysis are from pp collisions at $\s=13~\TeV$ collected in 2016, 2017, and 2018. A minimum-bias (MB) trigger was used, based on coincident signals in V0A and V0C. To enrich the data sample in the highest multiplicity regions, a high-multiplicity trigger based on a minimum threshold for the V0 amplitudes (HMV0) was used as well. The data sample collected with such a trigger corresponds to the 0.17\% highest-multiplicity events out of all inelastic collisions with at least one charged particle in the pseudorapidity range $|\eta|<1$ (denoted as INEL $>$ 0).
Offline selections were applied to remove background from beam--gas collisions, as described in Ref.~\cite{ALICE:2020swj}. Events with multiple reconstructed primary vertices were rejected. The remaining pile-up events were at a percent level and, therefore, did not affect the present analysis. Only the events with a primary vertex reconstructed within $|z_{\rm vtx}|<10$~cm from the nominal interaction point along the beam-line direction were considered for the analysis. To select events in the \inelgtrz class, at least one track segment reconstructed with the first two ITS layers (denoted as tracklet) within the pseudorapidity region $|\eta|<1$ was required. After these selections, the integrated luminosities are about $\Lint\approx32$  $\mathrm{nb}^{-1}$ for the MB triggered events, and $\Lint\approx7.7~\mathrm{pb}^{-1}$ for the HMV0 triggered events~\cite{ALICE:2021npz}. The event multiplicity was determined in the forward rapidity region, exploiting the sum of signal amplitudes in the V0A and V0C scintillators, V0M, and defining its percentile distribution, $p_\mathrm{V0M}$. Low $p_\mathrm{V0M}$ values represent high-multiplicity events. The definition of the mean multiplicity density ($\avdndeta_{|\eta|<0.5}$) of charged-primary particles at midrapidity is given in Ref.~\cite{ALICE:2021leo}. It was obtained by converting the measured event multiplicities as described in Ref.~\cite{ALICE:2020swj}. Table~\ref{tab:multbins} summarises the multiplicity event classes at forward rapidity used in this analysis ($p_\mathrm{V0M}(\%)$) and the corresponding values for $\avdndeta_{|\eta|<0.5}$, together with the corresponding value for the multiplicity-integrated class~\cite{ALICE:2020swj}.

Monte Carlo simulations were utilised in the analysis mainly for the machine-learning training, and to obtain the correction factors for the limited detector acceptance as well as the reconstruction and selection efficiencies. They were obtained by simulating pp collisions with the \pythiaprec event generator~\cite{Sjostrand:2006za, Sjostrand:2014zea} (Monash-13 tune~\cite{Skands:2014pea}). In order to enrich the simulated data samples of prompt and non-prompt D mesons, either a \ccbar or \bbbar quark pair was required in each simulated \textsc{PYTHIA} pp event and D mesons were forced to decay into the hadronic channels of interest for the analysis. The generated particles were transported through the apparatus by using the \geant package~\cite{Brun:1994aa}.

\begin{table}[!t]
  \centering
  \caption{Summary of the multiplicity event classes at forward rapidity expressed in percentiles of the V0M signal amplitude ($p_\mathrm{V0M} (\%)$). The average charged-particle densities $\avdndeta_{|\eta|<0.5}$ at midrapidity are shown, together with the value corresponding to the multiplicity-integrated class. Multiplicity intervals are measured in experimental data down to the 0--0.1\% percentile, corresponding to the highest-multiplicity interval.}
\renewcommand*{\arraystretch}{1.2}
  \begin{tabular}{l|>{\centering\arraybackslash}p{0.2\linewidth}}
  \toprule
   Multiplicity interval & $\avdndeta_{|\eta|<0.5}$  \\
  \midrule
$[30,100]\%$ & $4.41\pm 0.05$  \\
$[0.1,30]\%$ & $13.81\pm 0.14$  \\
$[0,0.1]\%$ & $31.53\pm 0.38$  \\
\midrule
  \inelgtrz & $6.93 \pm 0.09$ \\
\bottomrule
\end{tabular}
  \label{tab:multbins}
\end{table}

\section{Data analysis}
\label{sec:analysis}
The \Dzero and \Dplus mesons and their charge conjugates were reconstructed via the hadronic decay channels \DzerotoKpi, with branching ratio $\mathrm{BR} = (3.947 \pm 0.030) \%$, and \DplustoKpipi, with $\mathrm{BR} = (9.38 \pm 0.16) \%$~\cite{ParticleDataGroup:2022pth}. D-meson candidates were built by combining pairs or triplets of tracks with the proper charge signs, each track with $\pt > 0.3~\GeV/c$, $|y|>0.8$, at least 70 out of a maximum of 159 crossed TPC pad rows, a minimum number of two hits (out of six) in the ITS, with at least one in either of the two innermost layers, and a track fit quality $\chi^2/\rm{ndf} < 2$ in the TPC. These track-selection criteria reduce the D-meson acceptance in rapidity, which falls steeply to zero for $|y|>0.5$ at low \pt and for $|y|>0.8$ at $\pt>5~\GeV/c$. Thus, a fiducial acceptance selection $|y| < y_{\rm fid}(\pt)$, was applied to grant a uniform acceptance inside the rapidity range considered. The $y_{\mathrm{fid}}(\pt)$ value was defined as a second-order polynomial function, increasing from 0.5 to 0.8 in the transverse-momentum range $0 < \pt < 5~\GeV/c$, and as a constant term, $y_{\mathrm{fid}}=0.8$, for $\pt>5~\GeV/c$~\cite{ALICE:2021mgk}.

To suppress the large combinatorial background and to separate at the same time the contribution of prompt and non-prompt D mesons, a machine-learning approach with multi-class classification, based on Boosted Decision Trees (BDT) provided by the \textsc{XGBoost}~\cite{Chen:2016XST,barioglio_luca_2021_5070132} library was adopted. Signal samples of prompt and non-prompt D mesons for the BDT training were obtained from \pythia simulations as described in Sec.~\ref{sec:detector}. The background samples were obtained from candidates in the sideband region in the data, i.e.~in the interval $5\sigma < |\Delta M| < 9\sigma$ of the invariant mass distribution, where $\Delta M$ is the deviation between the invariant mass of
the candidate and the mean of a Gaussian function describing the signal peak and $\sigma$ is the Gaussian width. The training procedures are the same as reported in Ref.~\cite{ALICE:2021mgk}. Before the training, loose kinematic and topological selections were applied to the D-meson candidates.
The D-meson candidate information used for training the BDT models was mainly based on the displacement of the tracks from the primary vertex, the impact parameter of the D-meson daughter tracks, the distance between the D-meson decay vertex and the primary vertex, the cosine of the pointing angle between the D-meson candidate line of flight (the vector connecting the primary and secondary vertices) and its reconstructed momentum vector, and the PID information of the decay tracks. Independent BDTs were trained for each D-meson species and \pt interval in the multiplicity-integrated sample. Subsequently,  the BDTs were applied to the real data sample in which the type of candidate is unknown. The BDT outputs are related to the candidate probability to be a non-prompt D meson or combinatorial background. Selections on the BDT outputs were optimised to obtain a high non-prompt D-meson fraction while maintaining a reliable signal extraction (with statistical significance larger than 5).

\begin{figure*}[!tb]
\begin{center}
\includegraphics[width=0.85\textwidth]{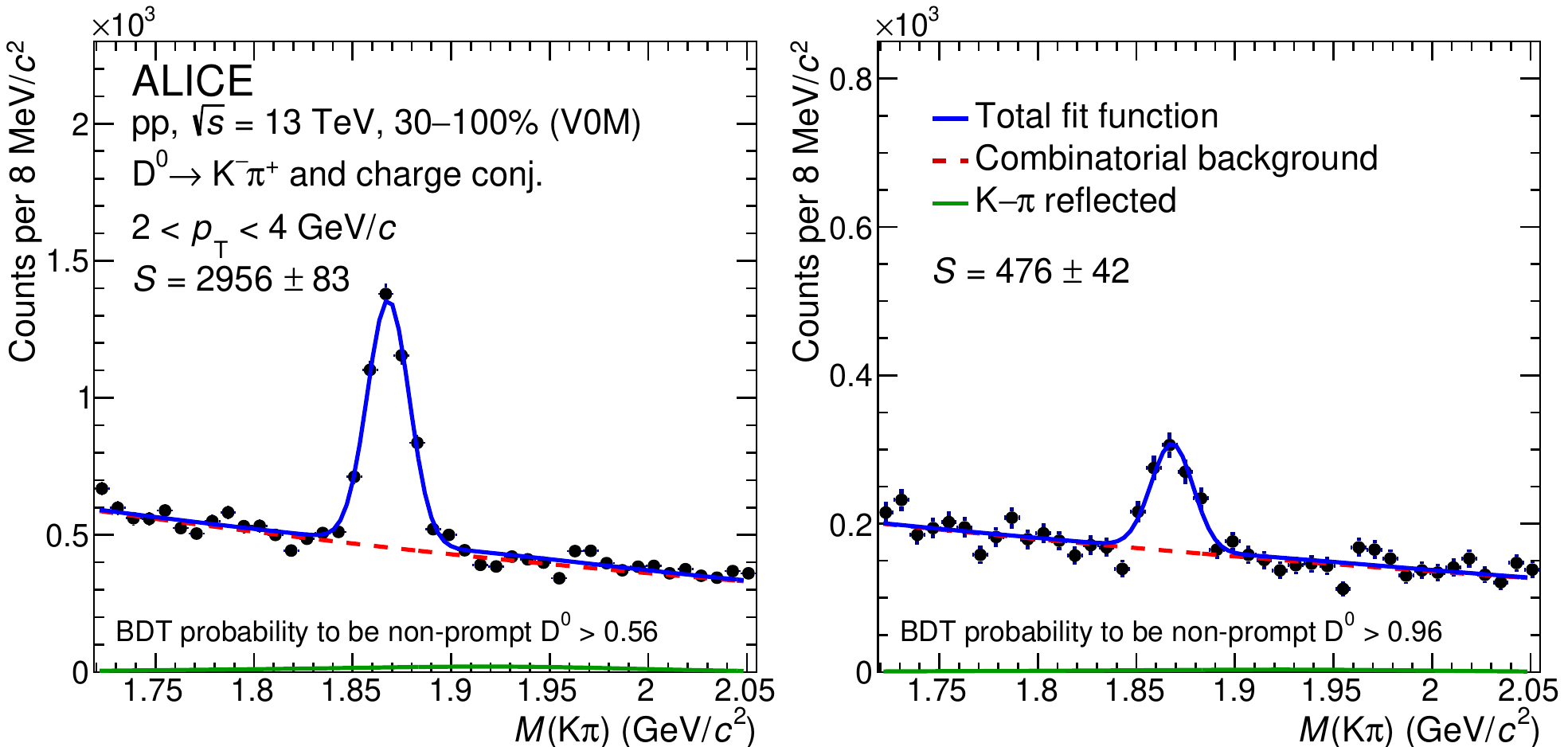}
\includegraphics[width=0.85\textwidth]{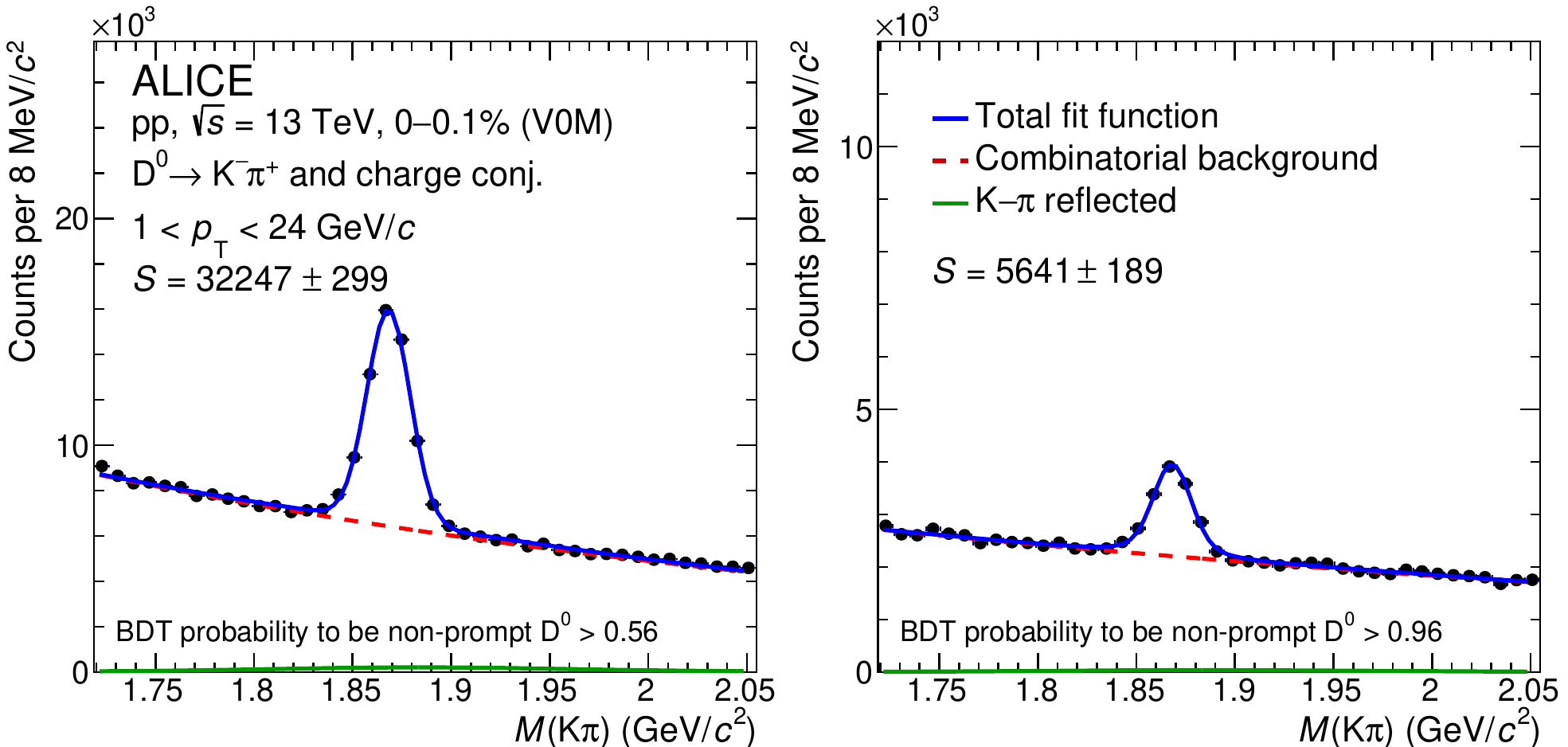}
\caption{Invariant-mass distribution of \Dzero candidates and their charge conjugates in selected \pt and multiplicity intervals. The blue solid curves show the total fit function and the red dashed curves show the combinatorial-background contribution. The green solid lines represent the reflection contribution. The raw-yield ($S$) values are reported together with their statistical uncertainties resulting from the fit. Top row: \Dzero mesons in the $2 < \pt < 4~\GeV/c$ interval for the low multiplicity class. Bottom row: \Dzero mesons in the $1 < \pt < 24~\GeV/c$ interval for the high multiplicity class. The corresponding BDT probability minimum threshold for the candidate selection is reported. The left (right) column corresponds to the prompt (non-prompt) \Dzero meson candidates dominated sample.} 
\label{fig:D0mass} 
\end{center}
\end{figure*}

\begin{figure*}[!tb]
\begin{center}
\includegraphics[width=0.85\textwidth]{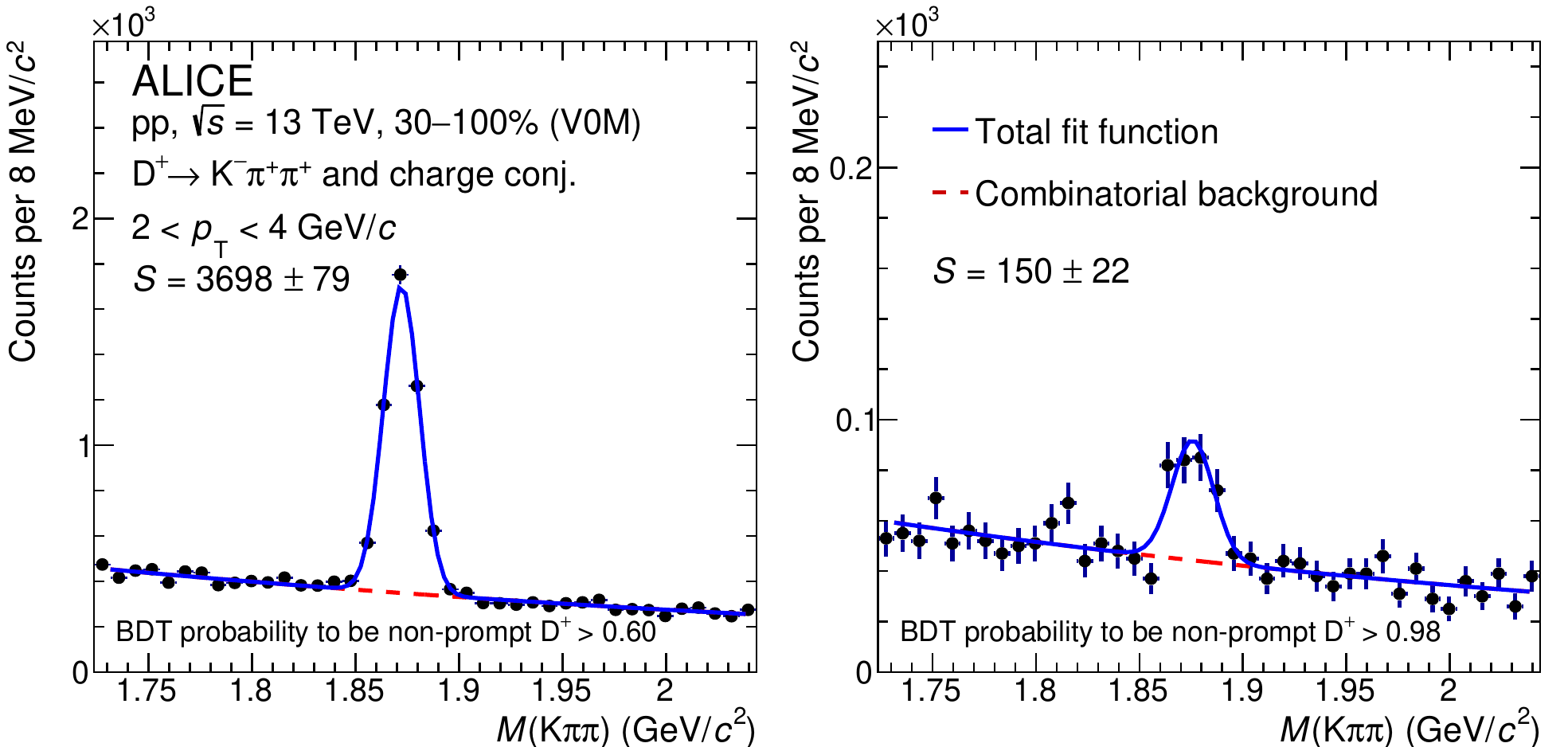}
\includegraphics[width=0.85\textwidth]{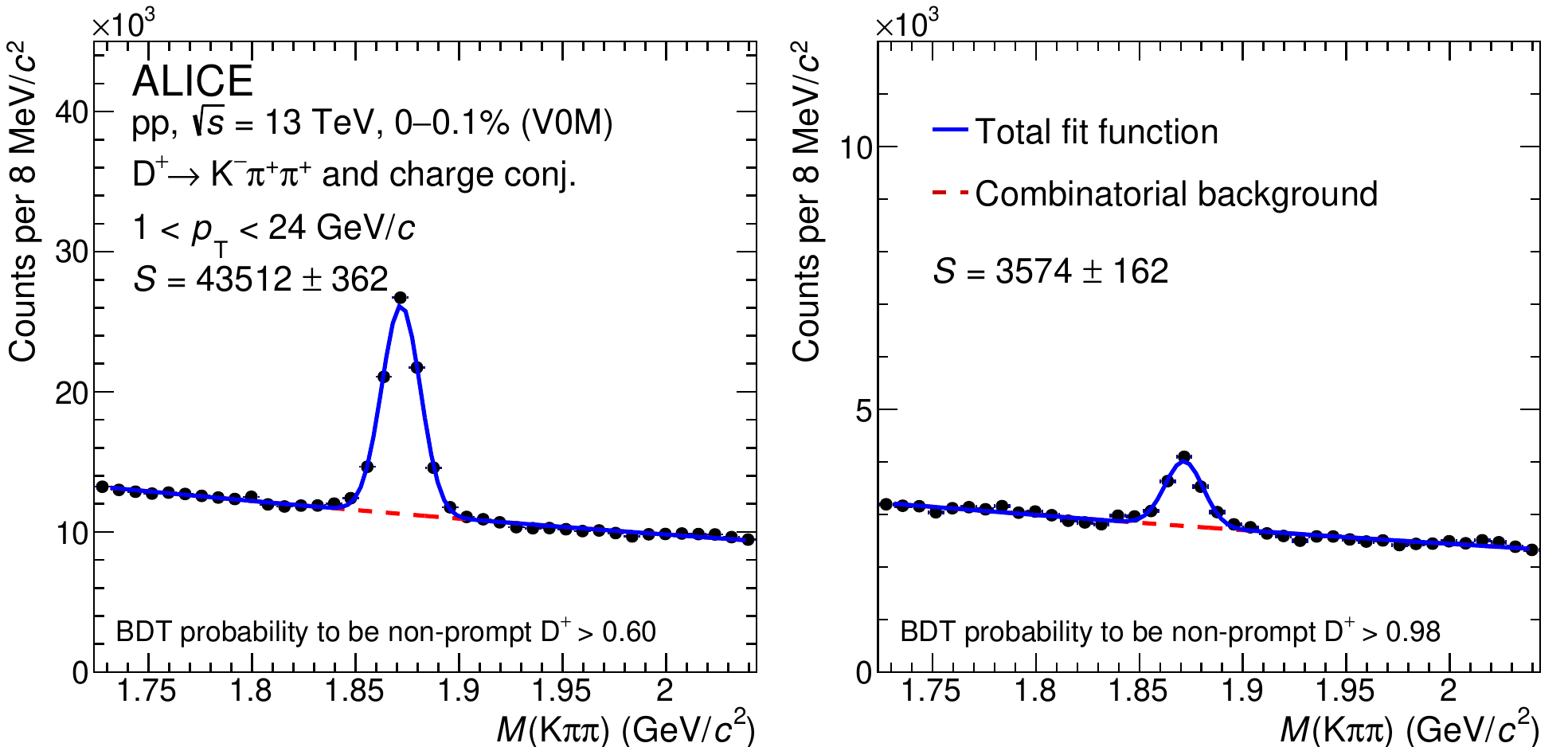}
\caption{Invariant-mass distribution of \Dplus candidates and their charge conjugates in selected \pt and multiplicity intervals. The blue solid curves show the total fit function and the red dashed curves show the combinatorial-background contribution. The raw-yield ($S$) values are reported together with their statistical uncertainties resulting from the fit.  Top row: \Dplus mesons in the $2 < \pt < 4~\GeV/c$ interval for the low multiplicity class. Bottom row: \Dplus mesons in the $1 < \pt < 24~\GeV/c$ interval for the high multiplicity class. The corresponding BDT probability minimum threshold for the candidate selection is reported. The left (right) column corresponds to the prompt (non-prompt) \Dplus meson candidates dominated sample. } 
\label{fig:Dplusmass} 
\end{center}
\end{figure*}

The signal extraction was performed in each \pt and multiplicity interval via a binned maximum-likelihood fit to the candidate invariant-mass distribution. The raw yields could be extracted in the transverse momentum interval $ 1 < \pt < 24~\GeV/c $ and in six subranges, for both \Dzero and \Dplus mesons. A Gaussian function and an exponential function were used to describe the signal peak and the background distribution, respectively. To improve the stability of the fits, the widths of the D-meson signal peaks were fixed to the values extracted from data samples dominated by prompt candidates, given the naturally higher abundance of prompt compared to non-prompt D mesons. In addition, for the \Dzero meson, the contribution of signal candidates to the invariant-mass distribution with the wrong decay-particle mass assignment (reflections) was included in the fit. It was parameterised by fitting the invariant-mass distribution of
reflections with a double Gaussian function, and normalised according to the reflection-to-signal ratio from the \pythia simulations. The contribution of reflections to the raw yield is about $0.5\%\mbox{--}4\%$, increasing with increasing \pt. Examples of invariant-mass fits with different contribution of signal from beauty-hadron decays in the $2 < \pt < 4~\GeV/c$ interval for the lowest multiplicity class and in the $1 < \pt < 24~\GeV/c$ interval for the highest multiplicity class are shown in Fig.~\ref{fig:D0mass} and Fig.~\ref{fig:Dplusmass} for \Dzero and \Dplus mesons, respectively. 
Based on the selections on the BDT outputs, samples dominated by non-prompt (prompt) candidates were selected by requiring low probability for a candidate to be combinatorial background and a high (low) probability to be non-prompt. The invariant-mass fits from non-prompt (prompt) enhanced samples are shown in each right (left) panel, indicating the corresponding selection applied on the BDT output score related to the probability to be a non-prompt D meson.

In each \pt and multiplicity interval, the fraction of non-prompt D mesons, \fnonprompt, was estimated by sampling the raw yield with different BDT selections related to the candidate probability of being a non-prompt D meson. In this way, a set of raw yields $\rawY{i}$ (index i refers to a given selection on the BDT output) with different contributions from  prompt and non-prompt D mesons was obtained. These raw yields are related to the corrected yields of prompt (\Np) and non-prompt (\Nnp) D mesons via the acceptance-times-efficiency (\eff) factors as

\begin{equation}
   \effP{i}\times \Np +  \effNP{i}\times \Nnp - \rawY{i} = \delta_\mathrm{i}.
\label{eq:eq_set}
\end{equation}

In the above equation, the $\delta_\mathrm{i}$ term represents a residual originating from the uncertainties on \rawY{i}, \effNP{i}, and \effP{i}. The \eff factors were obtained from MC simulations as described in Sec.~\ref{sec:detector}. They are different for prompt and non-prompt D mesons due to the different decay topology. Since the resolution of the reconstructed primary vertex depends on the multiplicity, the simulated events were weighted to reproduce the charged-particle multiplicity distribution measured in data for events that contain D-meson candidates having an invariant mass compatible with the one of the signal. After that, the \eff factors were computed for each BDT selection for prompt and non-prompt D mesons within the fiducial acceptance region. In the case of the number of sets of selections $n\geq 2$, a $\chi^2$ function can be defined based on the set of equations of Eq.~\ref{eq:eq_set}, which can be minimised to obtain \Np and \Nnp. More details can be found in Ref.~\cite{ALICE:2021mgk}. The \Nnp and \Np values can be used to calculate the corrected fraction of non-prompt D mesons as follows
\begin{equation}
\label{eq:fnpromptSystem}
  \fnonprompt = \frac{ \Nnp}{ \Nnp+ \Np}.
\end{equation}
In addition, the ratio between the fraction of non-prompt D mesons measured in each multiplicity interval and the one measured in the \inelgtrz class of events, \fnptofnpMB, was computed in multiplicity and \pt intervals in order to investigate the modification of the non-prompt fraction with respect to the one measured in the multiplicity-integrated sample.

\begin{figure*}[!tb]
\begin{center}
\includegraphics[width=0.48\textwidth]{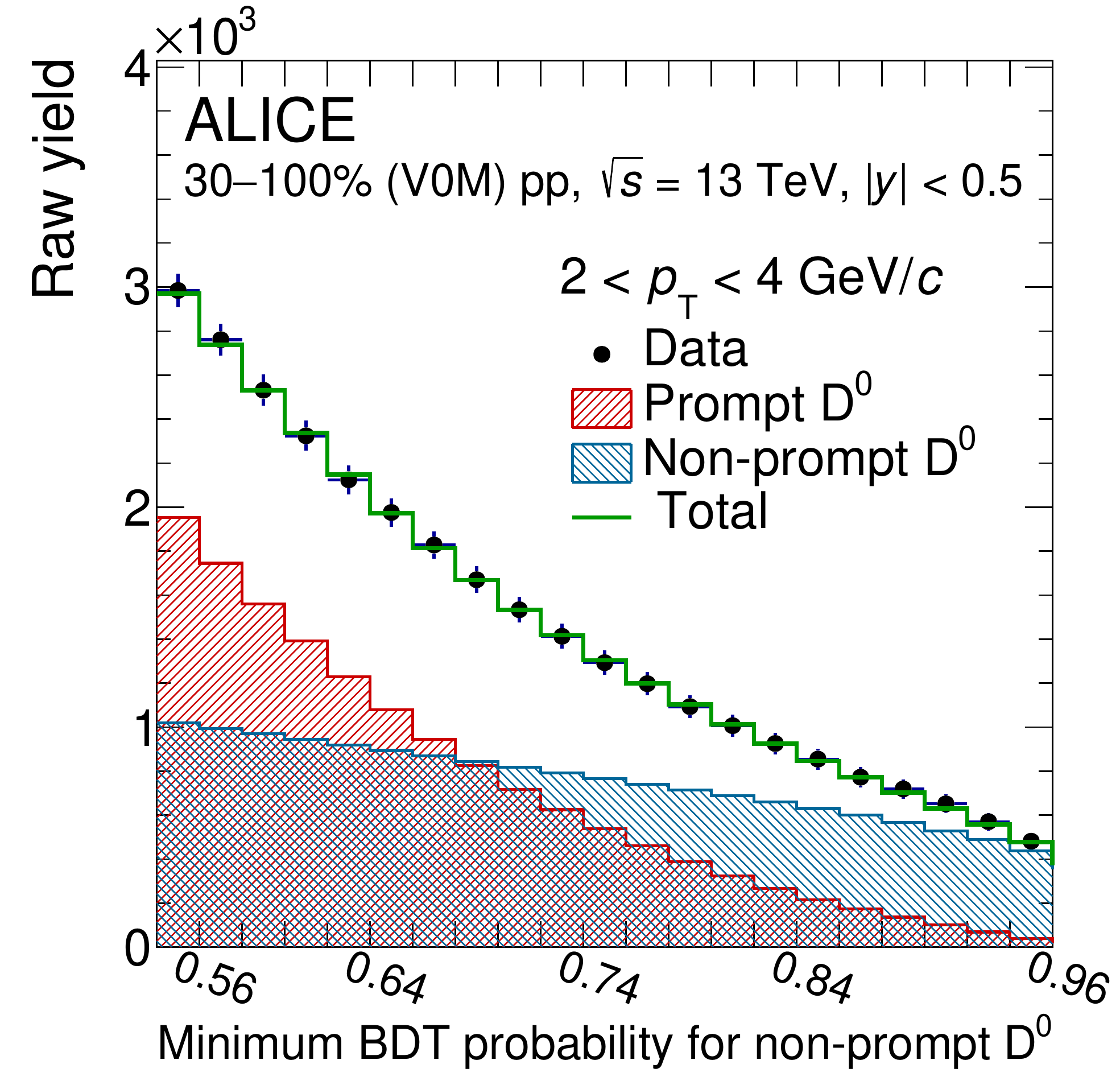}
\includegraphics[width=0.48\textwidth]{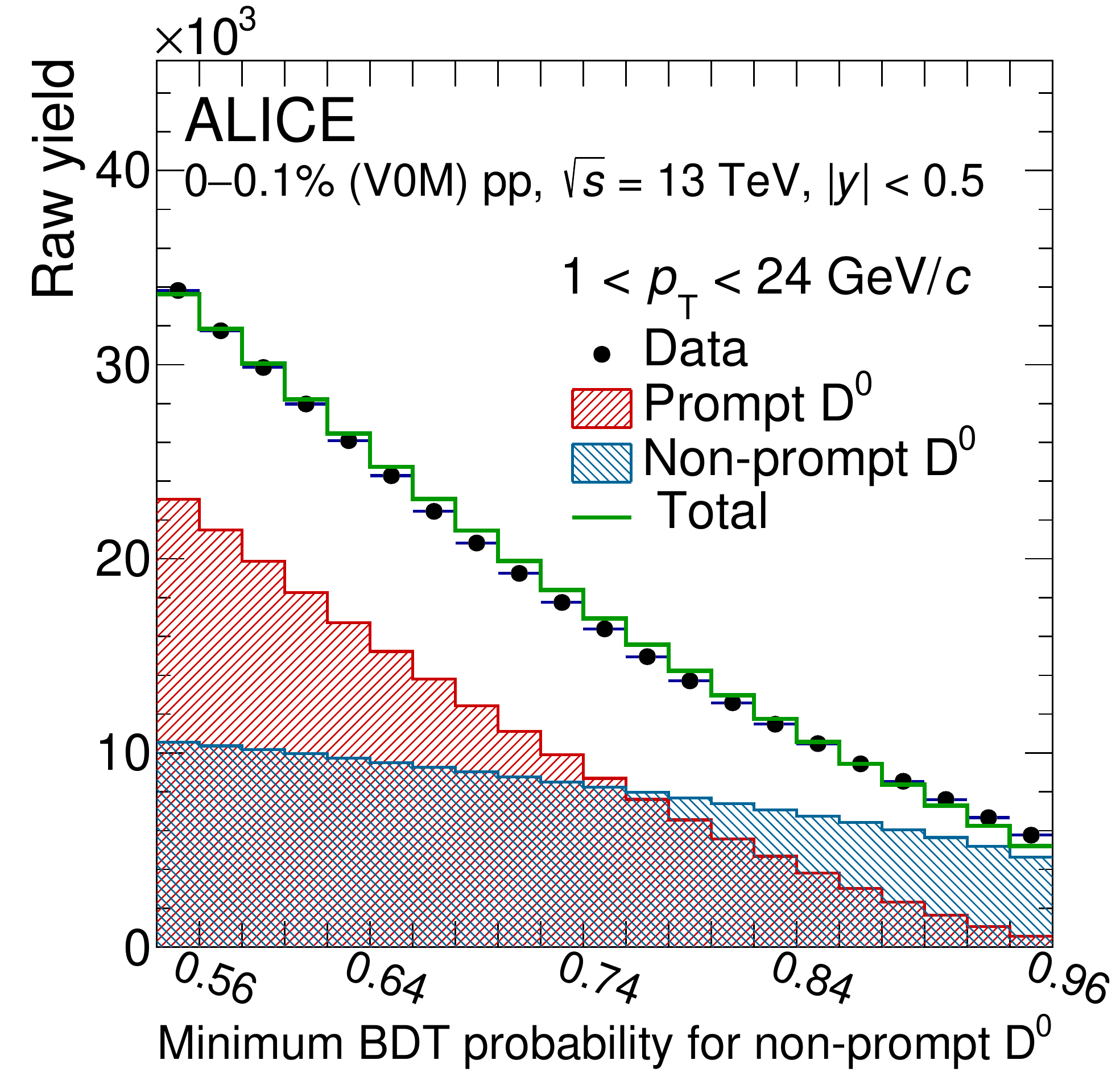}
\includegraphics[width=0.48\textwidth]{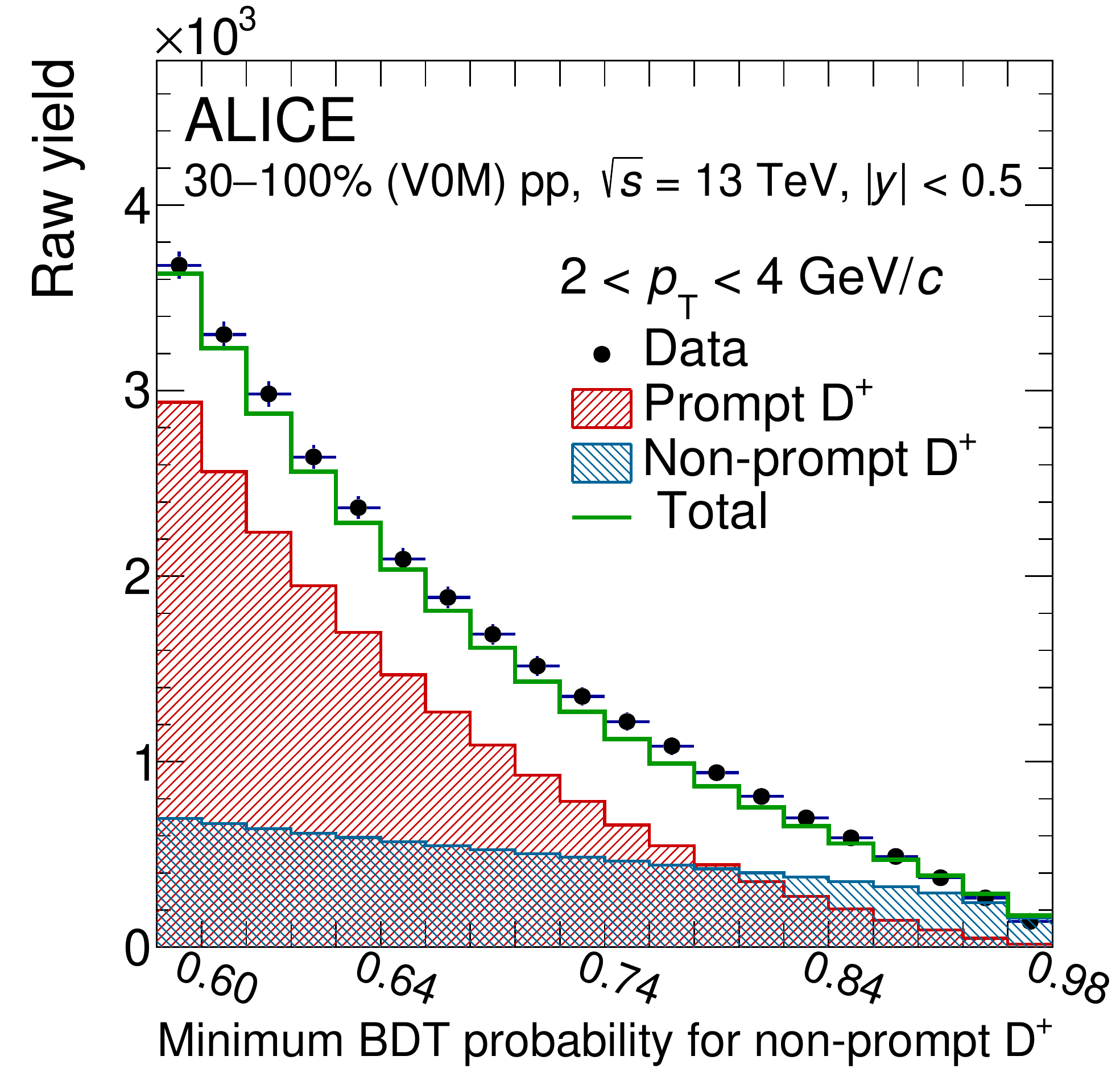}
\includegraphics[width=0.48\textwidth]{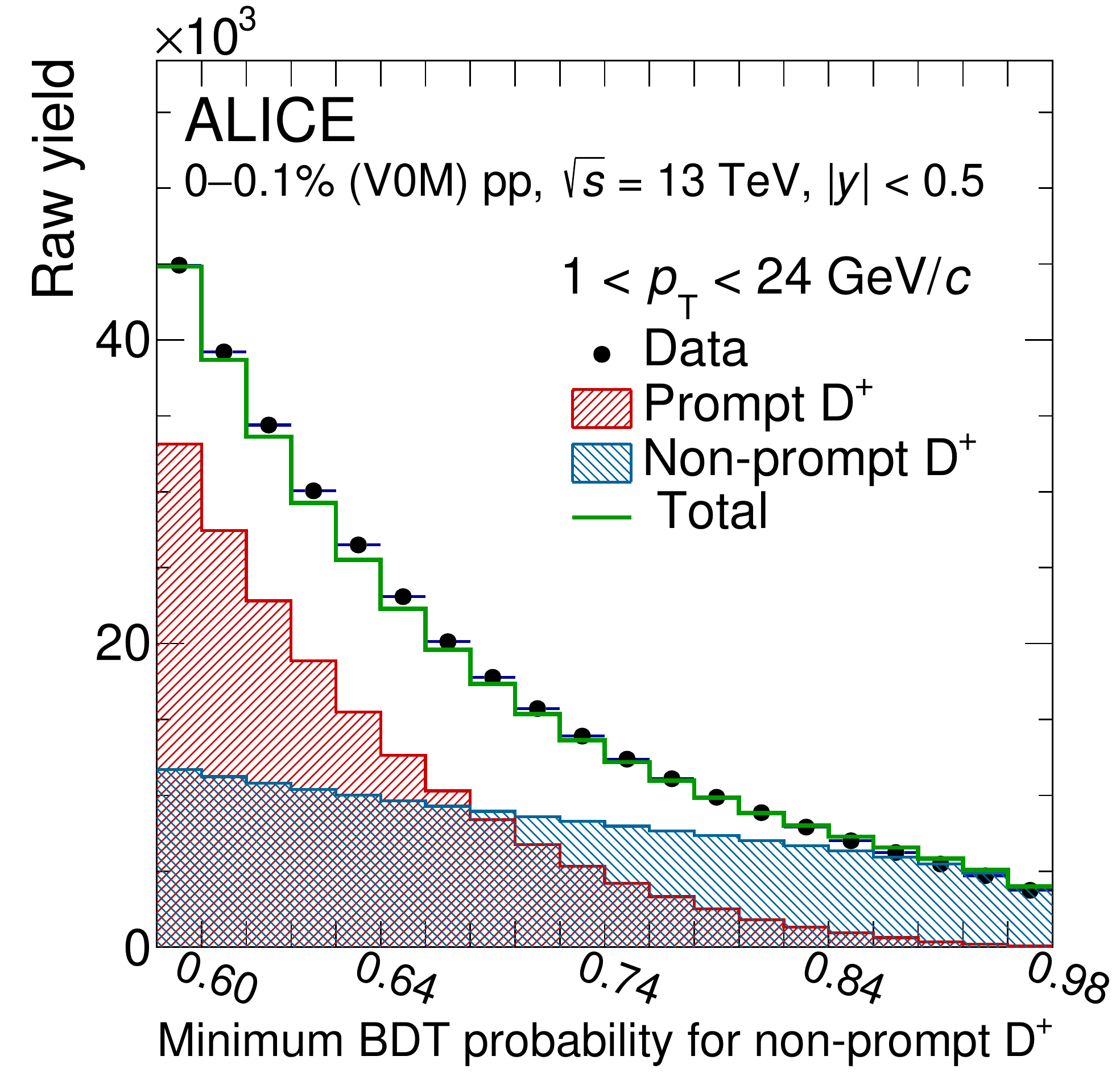}
\caption{Examples of raw-yield distribution as a function of the BDT-based selection employed in the $\chi^2$-minimisation procedure adopted for the determination of \fnonprompt of D mesons. Top row: \Dzero mesons in low multiplicity (left) and high multiplicity (right) classes. Bottom row: \Dplus mesons in low multiplicity (left) and high multiplicity (right) classes.} 
\label{fig:CutVariation} 
\end{center}
\end{figure*}

Figure~\ref{fig:CutVariation} shows an example of the raw-yield distribution as a function of the BDT-based selection used in the $\chi^2$-minimisation procedure for \Dzero (top panels) and \Dplus (bottom panels) mesons in the transverse-momentum intervals $2 < \pt < 4~\GeV/c$ and $1 < \pt < 24~\GeV/c$ for the low-multiplicity and high-multiplicity classes of events, respectively. The raw yield decreases with increasing minimum threshold for the probability to be a non-prompt D meson, corresponding to an increasing non-prompt D fraction. Note also that the raw yields used in this procedure are largely correlated among each other, implying that adjacent data points are expected to fluctuate in the same direction. The prompt and non-prompt components of the raw yields for each BDT-based selection obtained from the $\chi^2$-minimisation procedure as $\effP{i} \times \Np$ and $\effNP{i} \times\Nnp$, are reported as the red and blue distributions, and their sum is represented by the green histogram.

\section{Systematic uncertainties}
\label{sec:syst}

The values of systematic uncertainty on the non-prompt D-meson fraction were estimated with procedures similar to those described in Refs.~\cite{ALICE:2021mgk,ALICE:2021npz}. They include the uncertainties on (i) the raw-yield extraction from the invariant-mass distributions; 
(ii) the selection efficiency estimation;
(iii) the dependency of the efficiency on the charged-particle multiplicity;
 and (iv) the D-meson \pt shape in the simulation.
The estimated values of the systematic uncertainties for some representative \pt intervals of \Dzero and \Dplus mesons are summarised in Table~\ref{tab:sysunc_yieldtable}. 

\begin{table}[tb]
\caption{Summary of the relative systematic uncertainties on the non-prompt \Dzero-, \Dplus-meson fractions in various \pt and multiplicity intervals.}
\centering
\renewcommand*{\arraystretch}{1.2}
\begin{tabular}[t]{l|>{\centering}p{0.05\linewidth}>{\centering}p{0.05\linewidth}|>{\centering}p{0.05\linewidth}>{\centering}p{0.05\linewidth}|>{\centering}p{0.05\linewidth}>{\centering}p{0.05\linewidth}|>{\centering}p{0.05\linewidth}>{\centering}p{0.05\linewidth}|>{\centering}p{0.05\linewidth}>{\centering}p{0.05\linewidth}>{\centering}p{0.05\linewidth}>{\centering\arraybackslash}p{0.06\linewidth}}
\toprule
 Meson
 & \multicolumn{2}{c|}{\Dzero} 
 & \multicolumn{2}{c|}{\Dplus} 
 & \multicolumn{2}{c|}{\Dzero}
 & \multicolumn{2}{c}{\Dplus} 

 \\
$\pt~(\GeV/c)$  
& $2\mbox{--}4$   & \multicolumn{1}{c|}{$8\mbox{--}12$} 
& $2\mbox{--}4$   & \multicolumn{1}{c|}{$8\mbox{--}12$} 
& $1\mbox{--}2$   & \multicolumn{1}{c|}{$12\mbox{--}24$}  
& $1\mbox{--}2$  & \multicolumn{1}{c}{$12\mbox{--}24$}\\
\midrule
 & \multicolumn{4}{c|}{\fnonpromptLow} 
 & \multicolumn{4}{c}{\fnonpromptHig}\\
 \midrule
Raw-yield extraction                             
& 2\%   & \multicolumn{1}{c|}{2\%}   
& 3\%   & \multicolumn{1}{c|}{3\%}   
& 5\%   & \multicolumn{1}{c|}{2\%}  
& 6\% & \multicolumn{1}{c}{6\%}  \\
Efficiency estimation                          
& 2\%  & \multicolumn{1}{c|}{2\%}   
& 4\%   & \multicolumn{1}{c|}{5\%}   
& 6\%   & \multicolumn{1}{c|}{3\%}   
& 5\%  & \multicolumn{1}{c}{5\%}\\
\multicolumn{1}{l|}{MC multiplicity distribution}  
& 2\%  & \multicolumn{1}{c|}{1\%}   
& 4\%   & \multicolumn{1}{c|}{0\%}   
& 0\%   & \multicolumn{1}{c|}{0\%}   
& 0\%  & \multicolumn{1}{c}{0\%} \\
MC D-meson \pt distribution                           
& 6\%   & \multicolumn{1}{c|}{3\%}    
& 3\%   & \multicolumn{1}{c|}{1\%}  
& 9\%   & \multicolumn{1}{c|}{3\%}   
& 8\%  & \multicolumn{1}{c}{3\%} 
\\
\toprule
 & \multicolumn{4}{c|}{\fnpLowtofnpMB} 
 & \multicolumn{4}{c}{\fnpHigtofnpMB}\\
\midrule
 Raw-yield extraction                             
& 2\%   & \multicolumn{1}{c|}{2\%}   
& 4\%   & \multicolumn{1}{c|}{4\%}   
& 4\%   & \multicolumn{1}{c|}{2\%}  
& 7\% & \multicolumn{1}{c}{5\%}  \\
Efficiency estimation                          
& 2\%  & \multicolumn{1}{c|}{3\%}   
& 4\%   & \multicolumn{1}{c|}{5\%}   
& 4\%   & \multicolumn{1}{c|}{2\%}   
& 4\%  & \multicolumn{1}{c}{3\%}\\
\multicolumn{1}{l|}{MC multiplicity distribution}  
& 0\%  & \multicolumn{1}{c|}{2\%}   
& 0\%   & \multicolumn{1}{c|}{0\%}   
& 1\%   & \multicolumn{1}{c|}{0\%}   
& 4\%  & \multicolumn{1}{c}{0\%} \\
MC D-meson \pt distribution                           
& 4\%   & \multicolumn{1}{c|}{2\%}    
& 5\%   & \multicolumn{1}{c|}{1\%}  
& 0\%   & \multicolumn{1}{c|}{0\%}   
& 1\%  & \multicolumn{1}{c}{1\%}\\ 
 
\bottomrule
\end{tabular}
\label{tab:sysunc_yieldtable}	
\end{table}

The systematic uncertainty of the raw-yield extraction was evaluated by repeating the fits to the invariant-mass distribution varying the fit range and the functional form of the background and signal fit functions. To further test the sensitivity to the line shape of the signal, a bin-counting method, in which the signal yield
was obtained by integrating the background-subtracted invariant-mass distribution within the $\pm 3\sigma$ region relative to the peak position, was used. In the case of \Dzero mesons, an additional contribution due to signal reflections in the invariant-mass distribution was estimated by varying the normalisation and the shape of the templates used for the reflections in the invariant-mass fits. The systematic uncertainty was defined as the RMS of the distribution of the resulting \fnonprompt obtained from all these variations and ranges from 2\% to 6\% depending on the D-meson species, multiplicity, and \pt interval.

The systematic uncertainty of the selection-efficiency determination, arising from possible imperfections of the description of the decay topologies or the detector resolution in the simulation, was estimated by using alternative sets of BDT-output selections for the procedure described in Section~\ref{sec:analysis}. In particular, stricter and looser selections were tested, as well as different combinations of selections adopted to define the system of equations described in Eq.~\ref{eq:eq_set}. A systematic uncertainty ranging from 2\% to 6\% was assigned. 

To estimate the systematic uncertainty on the sensitivity of the efficiency on the charged-particle multiplicity, due to the multiplicity dependence of the primary-vertex reconstruction resolution, the distribution of the number of tracklets in the MC simulation for each V0M class of events was weighted using the one obtained in the real data considering events containing a D-meson candidate, without requiring the invariant-mass region selection. The resulting effect on the \fnonprompt estimation ranges from 0\% to 4\%.

The systematic uncertainty on the efficiency calculation due to a possible difference between the real and simulated D-meson transverse-momentum distributions was estimated by evaluating the efficiency after reweighting the \pt shape from the \pythia generator to match the one from \fonll calculations, in addition to the reweighting of the multiplicity distribution mentioned above. The weights were applied to the \pt distributions of prompt D mesons and to the parent beauty-hadron \pt distributions in case of non-prompt D mesons. The assigned uncertainty ranges from 1\% to 9\%.

The aforementioned sources of systematic uncertainty were assumed to be uncorrelated among each other. The total systematic uncertainty is defined as the square root of the quadratic sum of the estimated values in each \pt and multiplicity interval. In order to assess the correlation between the systematic uncertainties on \fnonprompt in the different multiplicity intervals with respect to the one in the \inelgtrz sample, the effect of the variations and the estimation of the uncertainties were directly evaluated on the ratio \fnptofnpMB.

\section{Results}
\label{sec:results}
\begin{figure*}[!t]
\begin{center}
\includegraphics[width=0.95\textwidth]{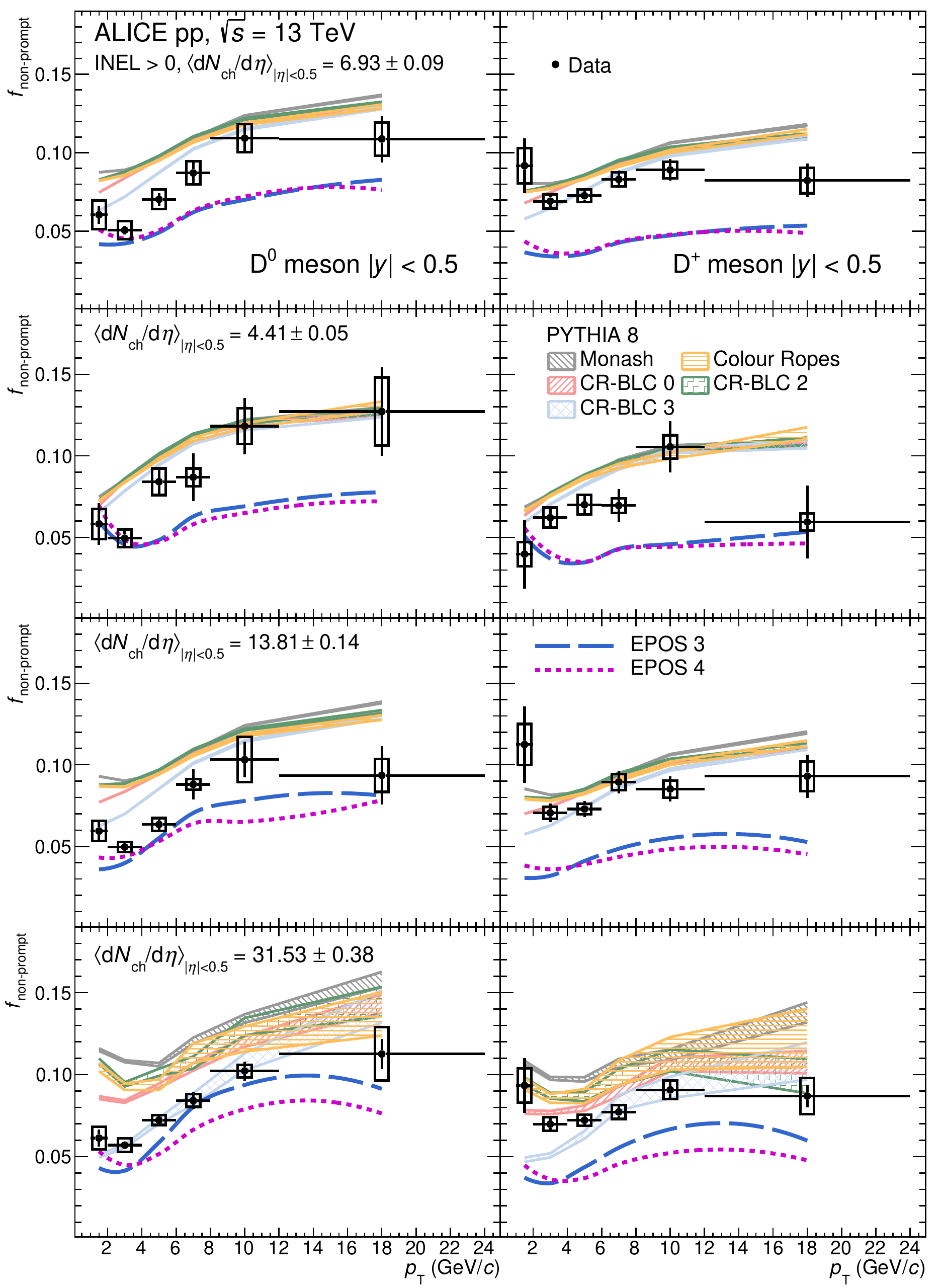}
\caption{Fractions of non-prompt \Dzero (left column) and \Dplus (right column) mesons as a function of \pt for the \inelgtrz class and the three multiplicity classes of events in pp collisions at $\s=13~\TeV$. The measurements are compared with the predictions obtained with \pythia~\cite{Christiansen:2015yqa} and \epos~\cite{Werner:2013tya} event generators.} 
\label{fig:D0DpFracsVsPt} 
\end{center}
\end{figure*}

The measured fractions of D-mesons originating from beauty-hadron decays, \fnonprompt, in pp collisions at $\s=13~\TeV$ are shown in Fig.~\ref{fig:D0DpFracsVsPt} as a function of \pt. The results are reported in different panels for \Dzero (left) and \Dplus (right) mesons and for the \inelgtrz class (top panels) and the three multiplicity classes of events (lower panels). The statistical and total systematic uncertainties are shown by vertical error bars and boxes, respectively.
In all the event classes and for both \Dzero and \Dplus mesons, \fnonprompt increases with \pt from 5\%--7\% to about 10\%. This increase is motivated by the harder \pt distribution of beauty hadrons compared to the charm ones, which is only partly compensated by the \BtoD decay kinematics~\cite{ALICE:2021mgk,Cacciari:2012ny}. The fraction of non-prompt \Dzero mesons is slightly larger than that of \Dplus mesons, as a consequence of the different branching ratios of B mesons with a \Dzero or \Dplus meson in the final state, and of the different charm-quark fragmentation fractions for the prompt D-meson production. This increasing trend is expected from pQCD calculations, as shown in Ref.~\cite{Bolzoni:2013vya}.
Measurements are compared to predictions from the \pythia~\cite{Sjostrand:2014zea, Bierlich:2022pfr} and \epos~\cite{Werner:2013tya,Werner:2023zvo} event generators. \pythia simulations were obtained using the standard Monash 2013 tune~\cite{Skands:2014pea} as well as with colour reconnection settings beyond-leading-colour approximation~\cite{Christiansen:2015yqa}, and with colour ropes~\cite{Bierlich:2022pfr} using \textsc{PYTHIA} version 8.307. Both the version 3.448 and 4.0.0 of the \epos MC generator were tested. In \eposfour, parallel partonic scatterings based on the $S$-matrix theory are implemented, leading to the factorisation of the hard and soft scales, particularly important for heavy quarks. This factorisation allows the computation of the PDFs within the EPOS framework itself. The \epos predictions presented in this paper do not include a hydrodynamic expansion of the system. However, the results were found not to significantly change if the latter is included. Following what was done for data, all \pythia and \epos simulations were selected according to percentiles of the \inelgtrz cross section based on the charged-particle multiplicity counts in the ALICE V0A and V0C acceptance. While all models qualitatively reproduce the increase of \fnonprompt with increasing \pt, \epos significantly underpredicts \fnonprompt of \Dplus mesons and \Dzero mesons in the \inelgtrz and in the lowest multiplicity classes of events, by up to a factor of two. Moreover, \eposthree predicts a slightly stronger multiplicity dependence compared to \eposfour. On the other hand, \pythia is generally closer to data but overpredicts \fnonprompt by approximately 20--30\%. No significant difference in the various \pythia settings tested in this work is observed, with the exception of the CR-BLC Mode 3 setting, which predicts a lower \Dplus and \Dzero non-prompt fraction especially in the two highest multiplicity intervals, providing a better description of the data.

\begin{figure*}[!t]
\begin{center}
\includegraphics[width=1.\textwidth]{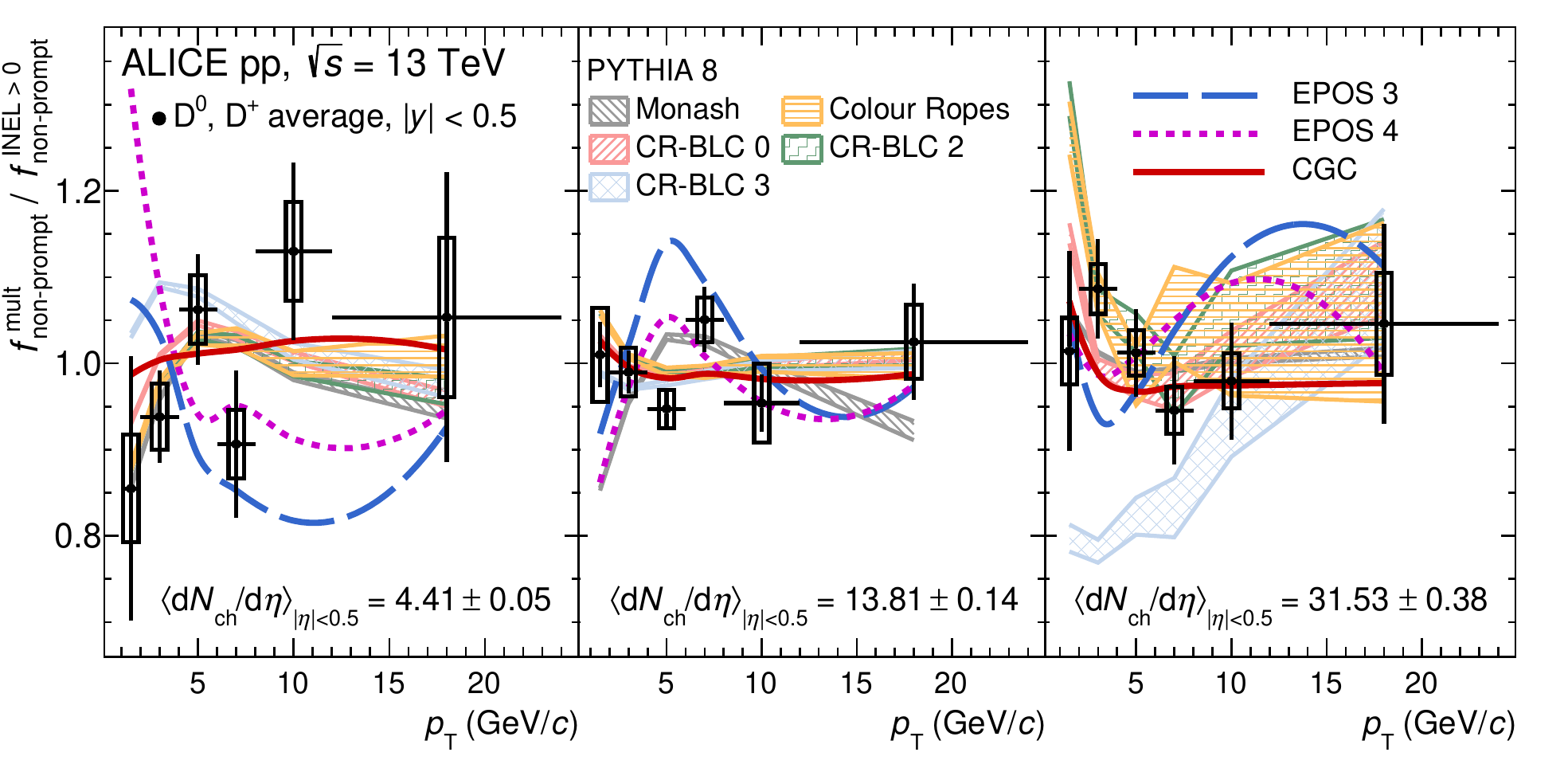}
\caption{Average fractions of non-prompt \Dzero and \Dplus mesons as a function of \pt for different multiplicity intervals normalised to the one measured in the \inelgtrz class of pp collisions at $\s=13~\TeV$. The measurements are compared with the predictions obtained with \pythia~\cite{Christiansen:2015yqa} and \epos~\cite{Werner:2013tya} event generators and the \cgc model.} 
\label{fig:D0DpFracRatiosVsPt} 
\end{center}
\end{figure*}

The ratio of the D-meson non-prompt fractions in the multiplicity classes relative to that in the \inelgtrz class, \fnptofnpMB, is shown in Fig.~\ref{fig:D0DpFracRatiosVsPt} as a function of transverse momentum for the three multiplicity classes. This double ratio isolates the relative variation of \fnonprompt as a function of the charged particle multiplicity from absolute scaling factors. The double ratio of \Dzero and \Dplus  was found to be compatible for all the multiplicity classes as expected. In order to improve the statistical precision, the average \Dzero and \Dplus \fnptofnpMB was computed. The average was computed using the inverse of the quadratic sum of the relative statistical and uncorrelated systematic uncertainties as weights. The systematic uncertainties were propagated through the averaging procedure considering the contributions from the raw-yield extraction and the selection efficiency as uncorrelated, while the other sources as fully correlated between the two D-meson species. In all multiplicity classes, the measured ratio is compatible with unity within uncertainties. This finding suggests similar production mechanisms of charm and beauty quarks as a function of multiplicity. The expectation obtained with \eposthree shows a modification of the \pt spectrum different for charm and beauty hadrons due to their different mass, which is not supported by the measurement. A qualitatively similar behaviour is obtained with \eposfour, which is more in agreement with the data, except for $\pt<4~\GeV/c$ in low-multiplicity events. All the \pythia configurations reproduce the measurements within the uncertainties, indicating a small influence of the hadronisation in the multiplicity dependence, except for the CR-BLC Mode 3 setting, which underestimates the data at low \pt in the high-multiplicity class of events. The data points are further compared to a \cgc model that includes the three-pomeron exchange mechanism~\cite{Schmidt:2020fgn}. In this model, the transition from the beauty quark to the charm hadron is modelled in a single step using \fbtoHc fragmentation functions measured in \ee collisions~\cite{Kneesch:2007ey}. Even though these fragmentation functions were shown to be unable to reproduce the measured cross sections of non-prompt D mesons in previous studies~\cite{ALICE:2021mgk}, they cancel in the \fnptofnpMB ratio and for this observable the \cgc predictions are consistent with the data within uncertainties.

\begin{figure*}[!t]
\begin{center}
\includegraphics[width=0.95\textwidth]{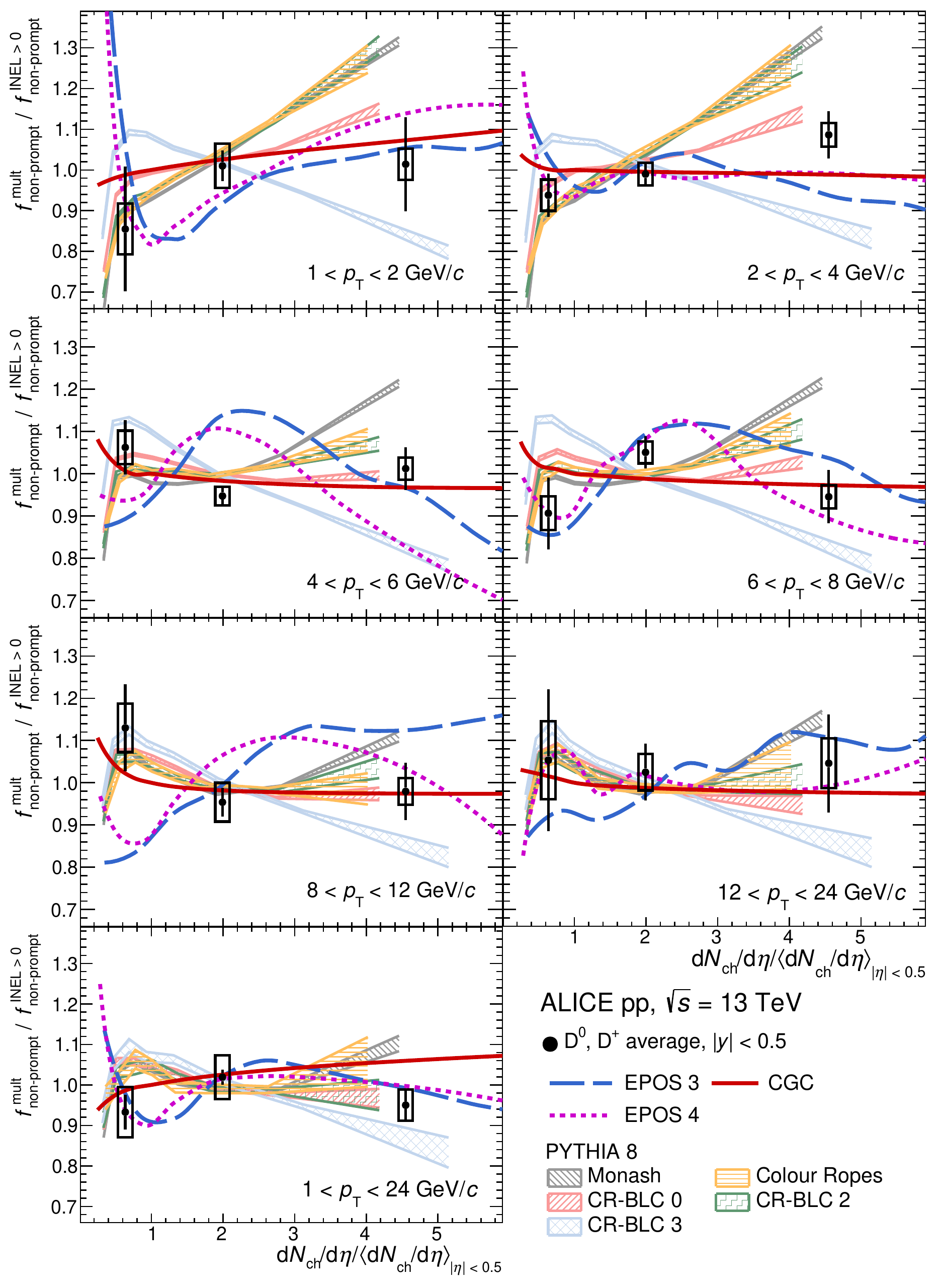}
\caption{Average fractions of non-prompt \Dzero and \Dplus mesons as a function of multiplicity, both normalised to the value corresponding to the \inelgtrz class, for pp collisions at $\s=13~\TeV$ in different \pt intervals and integrated in $1<\pt<24~\GeV/c$. The measurements are compared with predictions obtained with the \pythia~\cite{Christiansen:2015yqa} and \epos~\cite{Werner:2013tya} event generators and the \cgc model~\cite{Schmidt:2020fgn}.} 
\label{fig:D0DpFracRatiosVsMult} 
\end{center}
\end{figure*}

\begin{figure*}[!t]
\begin{center}
\includegraphics[width=1.\textwidth]{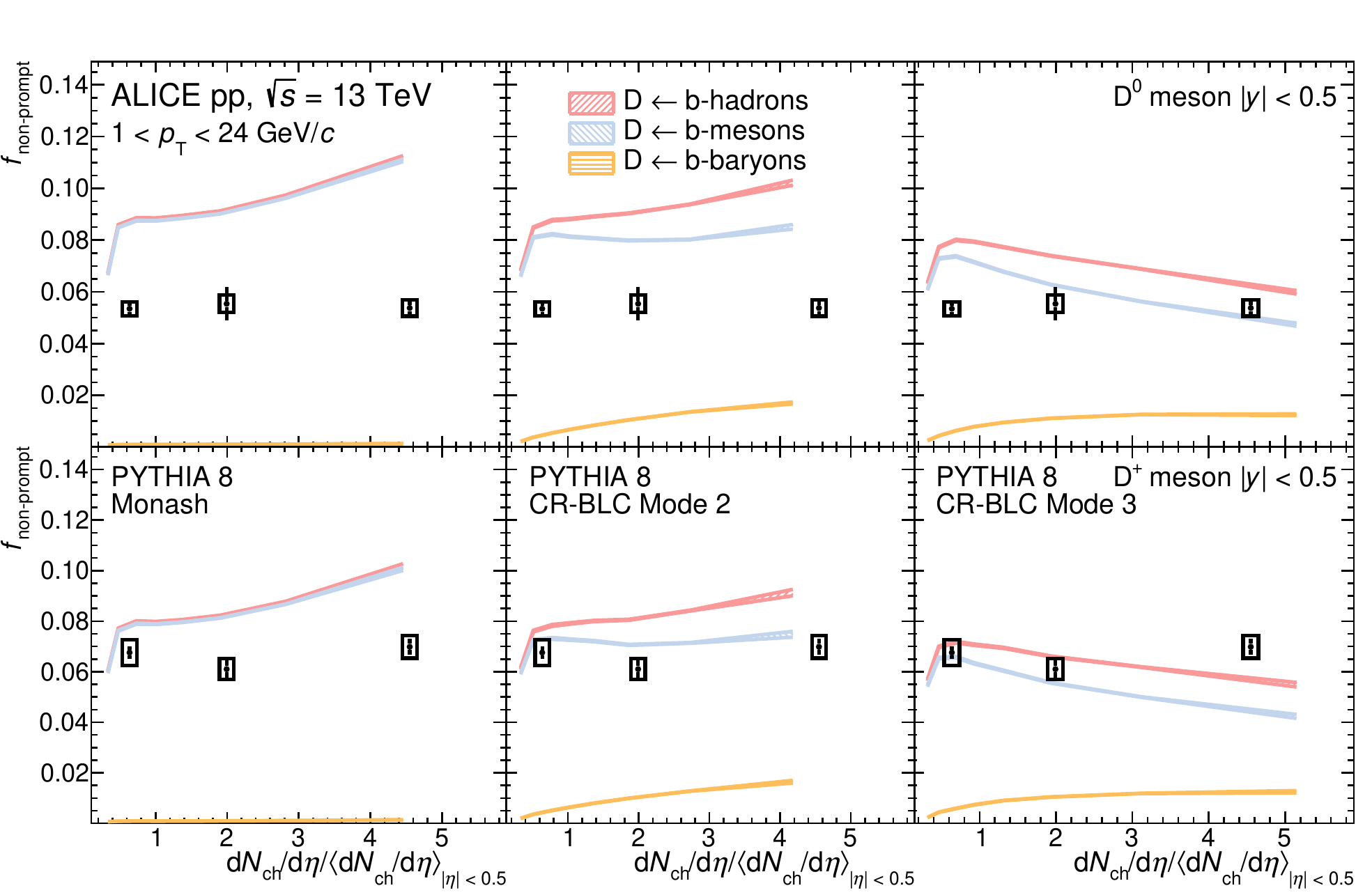}
\caption{Fractions of non-prompt \Dzero (first row) and \Dplus (second row) mesons in $1<\pt<24~\GeV/c$ as a function of multiplicity for pp collisions at $\s=13~\TeV$ compared with predictions obtained with the \pythia~\cite{Christiansen:2015yqa} event generator. The contributions from beauty meson and baryon decays in \pythia are displayed separately.} 
\label{fig:D0DpFracRatiosVsMultBreakdown} 
\end{center}
\end{figure*}

\begin{figure*}[!t]
\begin{center}
\includegraphics[width=0.95\textwidth]{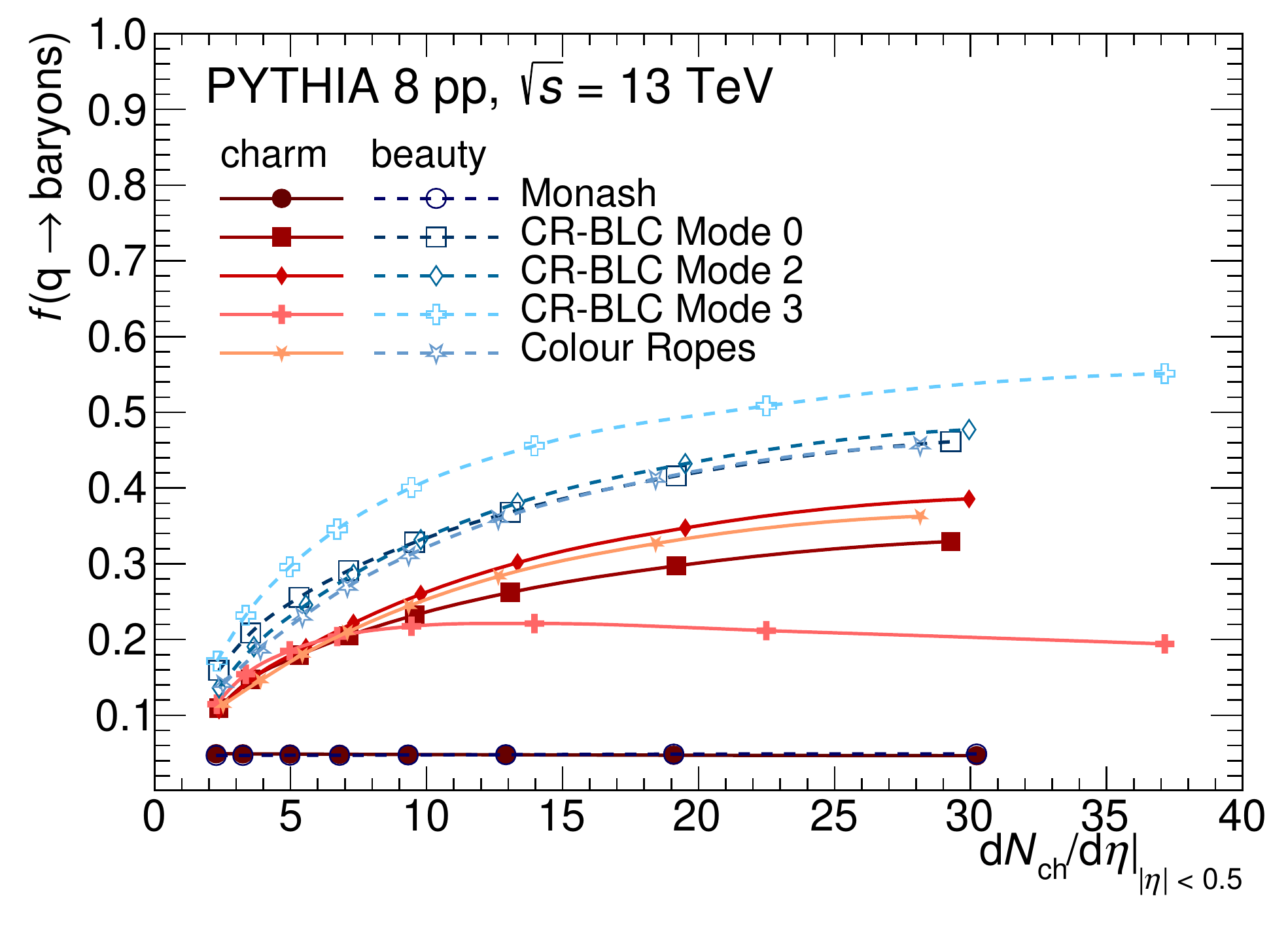}
\caption{Fraction of charm and beauty quarks hadronising to baryons as a function of the charged particle multiplicity at midrapidity in \pythia~\cite{Christiansen:2015yqa} simulations with different tunes.} 
\label{fig:BaryonFraction} 
\end{center}
\end{figure*}

The specific multiplicity dependence of \fnptofnpMB can be studied in more detail by plotting the values obtained in each individual transverse 
momentum interval as a function of the charged-particle multiplicity density normalised to the value corresponding to the \inelgtrz class of events, as shown in Fig.~\ref{fig:D0DpFracRatiosVsMult}. In all \pt intervals, the average \Dzero and \Dplus \fnptofnpMB ratio is found to be compatible with unity, indicating a weak (if any) dependence of \fnonprompt with the charged-particle multiplicity. Comparisons
with models reveal that the \epos event generator predicts a multiplicity dependence at intermediate transverse momentum ($4<\pt<6~\GeV/c$) which is ruled out by the data. At low \pt and multiplicity it predicts a rise of \fnonprompt which is also not supported by the data. At lower and higher \pt in the other multiplicity intervals, instead, \epos predicts a milder charged-particle multiplicity dependence and is hence closer to the data. Moreover, the multiplicity-independence of \cgc predictions is also consistent with the data. Finally, most \pythia predictions are consistent with the data, with the notable exception of CR-BLC Mode 3 results, in which the double ratio is shown to decrease with multiplicity. This behaviour can be further investigated by isolating the double ratio for D mesons originating from beauty-meson and beauty-baryon decays in each of the specific \pythia configurations being used, as represented in Fig.~\ref{fig:D0DpFracRatiosVsMultBreakdown}. While in all cases the \fnptofnpMB ratio from beauty baryons increases systematically with multiplicity, the Mode 3 setting results
in a decrease of this double ratio for D mesons originating from B-meson decays. More specifically, a clean MC-only test can be performed with the beyond-leading-colour tunes by calculating
the ratio of baryons and mesons at hadronisation time in \pythia as a function of multiplicity in each model,
as depicted in Fig.~\ref{fig:BaryonFraction}. Notably, CR-BLC Mode 3 differs from other \pythia predictions
due to the fact that, in that case, beauty quarks produce significantly more baryons, and charm quarks produce
fewer baryons than in other cases.
Consequently, the fraction of non-prompt D mesons decreases with the multiplicity as a combination of two effects. On the one side, charm quarks hadronise more to D mesons, increasing the prompt contribution to the D-meson production and, on the other side, beauty quarks will tend towards being contained in baryons, which in turn will feed preferentially into charm baryons such as the \LambdaC baryon. This strong preference towards beauty baryons is not favoured by current ALICE data, which essentially rules out the CR-BLC Mode 3 dynamics in favour of models in which \fnonprompt tends to either remain constant or increase slightly with multiplicity. Future studies of meson and baryon production in the beauty sector as a function of the charged-particle multiplicity will allow for firmer conclusions.

\section{Summary}
\label{sec:summary}

The fractions of the \Dzero and \Dplus mesons originating from beauty-hadron decays, \fnonprompt, were measured at midrapidity ($|y|<0.5$) in pp collisions at $\s=13~\TeV$ in events with at least a charged particle at midrapidity (\inelgtrz class of events) and as a function of charged-particle multiplicity and transverse momentum. Events with different charged-particle multiplicities were selected as percentiles of the \inelgtrz cross section based on the charged-particle multiplicity counts in the ALICE V0A and V0C at forward and backward rapidity.
The \Dplus and \Dzero \fnonprompt were observed to slightly increase from about 5\%--7\% for $1<\pt<3~\GeV/c$ to about 10\% for $8<\pt<24~\GeV/c$. The ratios \fnptofnpMB are compatible with unity both as a function of \pt and charged-particle multiplicity, suggesting either no or only a mild multiplicity dependence. This finding suggests a similar production mechanism of charm and beauty quarks as a function of multiplicity.

The measured \fnonprompt values are compared to predictions obtained with different MC generators. The \epos 3 and \epos 4 generators tend to underestimate the measurements, while \pythia with different tunes, including the colour reconnection mechanism beyond leading colour approximation and colour ropes, slightly overestimates the data. The variation of \fnonprompt with multiplicity is satisfactorily described by the MC simulations except for the $4<\pt<6~\GeV/c$ interval, where the \epos generator predicts a significant increase. In all the considered \pt intervals, the CR-BLC Mode 3 tune of \pythia foresees a decrease at high multiplicity. In that tune, this decrease with increasing multiplicity is motivated by an interplay between an increased fraction of charm quarks hadronising into mesons and an increased fraction of beauty quarks hadronising into baryons and is not favoured by data. Despite the fragmentation functions adopted prevented to reproduce the measured cross sections of non-prompt D mesons in previous studies~\cite{ALICE:2021mgk}, the ratio \fnptofnpMB is also described well by the \cgc model.
The comparison between data and theory models suggests a similar multiplicity dependence of charm- and beauty-hadron production and in particular, a different evolution of the baryon-to-meson ratio in the charm and beauty sectors is disfavoured.

The measurements presented in this paper provide an important test for production and hadronisation models in the charm and beauty sectors, and they pave the way for future studies of beauty-hadron production in pp collisions as a function of the charged-particle multiplicity.

%

\newenvironment{acknowledgement}{\relax}{\relax}
\begin{acknowledgement}
\section*{Acknowledgements}

The ALICE Collaboration would like to thank all its engineers and technicians for their invaluable contributions to the construction of the experiment and the CERN accelerator teams for the outstanding performance of the LHC complex.
The ALICE Collaboration gratefully acknowledges the resources and support provided by all Grid centres and the Worldwide LHC Computing Grid (WLCG) collaboration.
The ALICE Collaboration acknowledges the following funding agencies for their support in building and running the ALICE detector:
A. I. Alikhanyan National Science Laboratory (Yerevan Physics Institute) Foundation (ANSL), State Committee of Science and World Federation of Scientists (WFS), Armenia;
Austrian Academy of Sciences, Austrian Science Fund (FWF): [M 2467-N36] and Nationalstiftung f\"{u}r Forschung, Technologie und Entwicklung, Austria;
Ministry of Communications and High Technologies, National Nuclear Research Center, Azerbaijan;
Conselho Nacional de Desenvolvimento Cient\'{\i}fico e Tecnol\'{o}gico (CNPq), Financiadora de Estudos e Projetos (Finep), Funda\c{c}\~{a}o de Amparo \`{a} Pesquisa do Estado de S\~{a}o Paulo (FAPESP) and Universidade Federal do Rio Grande do Sul (UFRGS), Brazil;
Bulgarian Ministry of Education and Science, within the National Roadmap for Research Infrastructures 2020-2027 (object CERN), Bulgaria;
Ministry of Education of China (MOEC) , Ministry of Science \& Technology of China (MSTC) and National Natural Science Foundation of China (NSFC), China;
Ministry of Science and Education and Croatian Science Foundation, Croatia;
Centro de Aplicaciones Tecnol\'{o}gicas y Desarrollo Nuclear (CEADEN), Cubaenerg\'{\i}a, Cuba;
Ministry of Education, Youth and Sports of the Czech Republic, Czech Republic;
The Danish Council for Independent Research | Natural Sciences, the VILLUM FONDEN and Danish National Research Foundation (DNRF), Denmark;
Helsinki Institute of Physics (HIP), Finland;
Commissariat \`{a} l'Energie Atomique (CEA) and Institut National de Physique Nucl\'{e}aire et de Physique des Particules (IN2P3) and Centre National de la Recherche Scientifique (CNRS), France;
Bundesministerium f\"{u}r Bildung und Forschung (BMBF) and GSI Helmholtzzentrum f\"{u}r Schwerionenforschung GmbH, Germany;
General Secretariat for Research and Technology, Ministry of Education, Research and Religions, Greece;
National Research, Development and Innovation Office, Hungary;
Department of Atomic Energy Government of India (DAE), Department of Science and Technology, Government of India (DST), University Grants Commission, Government of India (UGC) and Council of Scientific and Industrial Research (CSIR), India;
National Research and Innovation Agency - BRIN, Indonesia;
Istituto Nazionale di Fisica Nucleare (INFN), Italy;
Japanese Ministry of Education, Culture, Sports, Science and Technology (MEXT) and Japan Society for the Promotion of Science (JSPS) KAKENHI, Japan;
Consejo Nacional de Ciencia (CONACYT) y Tecnolog\'{i}a, through Fondo de Cooperaci\'{o}n Internacional en Ciencia y Tecnolog\'{i}a (FONCICYT) and Direcci\'{o}n General de Asuntos del Personal Academico (DGAPA), Mexico;
Nederlandse Organisatie voor Wetenschappelijk Onderzoek (NWO), Netherlands;
The Research Council of Norway, Norway;
Commission on Science and Technology for Sustainable Development in the South (COMSATS), Pakistan;
Pontificia Universidad Cat\'{o}lica del Per\'{u}, Peru;
Ministry of Education and Science, National Science Centre and WUT ID-UB, Poland;
Korea Institute of Science and Technology Information and National Research Foundation of Korea (NRF), Republic of Korea;
Ministry of Education and Scientific Research, Institute of Atomic Physics, Ministry of Research and Innovation and Institute of Atomic Physics and University Politehnica of Bucharest, Romania;
Ministry of Education, Science, Research and Sport of the Slovak Republic, Slovakia;
National Research Foundation of South Africa, South Africa;
Swedish Research Council (VR) and Knut \& Alice Wallenberg Foundation (KAW), Sweden;
European Organization for Nuclear Research, Switzerland;
Suranaree University of Technology (SUT), National Science and Technology Development Agency (NSTDA), Thailand Science Research and Innovation (TSRI) and National Science, Research and Innovation Fund (NSRF), Thailand;
Turkish Energy, Nuclear and Mineral Research Agency (TENMAK), Turkey;
National Academy of  Sciences of Ukraine, Ukraine;
Science and Technology Facilities Council (STFC), United Kingdom;
National Science Foundation of the United States of America (NSF) and United States Department of Energy, Office of Nuclear Physics (DOE NP), United States of America.
In addition, individual groups or members have received support from:
European Research Council, Strong 2020 - Horizon 2020, Marie Sk\l{}odowska Curie (grant nos. 950692, 824093, 896850), European Union;
Academy of Finland (Center of Excellence in Quark Matter) (grant nos. 346327, 346328), Finland;
Programa de Apoyos para la Superaci\'{o}n del Personal Acad\'{e}mico, UNAM, Mexico.

\end{acknowledgement}

\bibliographystyle{utphys}   
\bibliography{bibliography}

\providecommand{\href}[2]{#2}\begingroup\raggedright\begin{thebibliography}{10}

\bibitem{ALICE:2019nxm}
{\bfseries ALICE} Collaboration, S.~Acharya {\em et~al.}, ``{Measurement of
  ${{\mathrm{D}}^0}$ , ${{\mathrm{D}}^+}$ , ${{\mathrm{D}}^{*+}}$ and
  ${{\mathrm{D}}^+_{\mathrm{s}}}$ production in pp collisions at
  ${\sqrt{{\textit{s}}}~=~5.02~{\text {TeV}}}$ with ALICE}'',
  \href{http://dx.doi.org/10.1140/epjc/s10052-019-6873-6}{{\em Eur. Phys. J. C}
  {\bfseries 79} (2019) 388}, \href{http://arxiv.org/abs/1901.07979}{{\ttfamily
  arXiv:1901.07979 [nucl-ex]}}.

\bibitem{ALICE:2021mgk}
{\bfseries ALICE} Collaboration, S.~Acharya {\em et~al.}, ``{Measurement of
  beauty and charm production in pp collisions at $ \sqrt{s} $ = 5.02 TeV via
  non-prompt and prompt D mesons}'',
  \href{http://dx.doi.org/10.1007/JHEP05(2021)220}{{\em JHEP} {\bfseries 05}
  (2021) 220}, \href{http://arxiv.org/abs/2102.13601}{{\ttfamily
  arXiv:2102.13601 [nucl-ex]}}.

\bibitem{ALICE:2021edd}
{\bfseries ALICE} Collaboration, S.~Acharya {\em et~al.}, ``{Prompt and
  non-prompt J/\ensuremath{\psi} production cross sections at midrapidity in
  proton-proton collisions at $ \sqrt{\mathrm{s}} $ = 5.02 and 13 TeV}'',
  \href{http://dx.doi.org/10.1007/JHEP03(2022)190}{{\em JHEP} {\bfseries 03}
  (2022) 190}, \href{http://arxiv.org/abs/2108.02523}{{\ttfamily
  arXiv:2108.02523 [nucl-ex]}}.

\bibitem{ALICE:2019nuy}
{\bfseries ALICE} Collaboration, S.~Acharya {\em et~al.}, ``{Measurement of
  electrons from semileptonic heavy-flavour hadron decays at midrapidity in pp
  and Pb--Pb collisions at $\sqrt{s_{\rm{NN}}}$ = 5.02 TeV}'',
  \href{http://dx.doi.org/10.1016/j.physletb.2020.135377}{{\em Phys. Lett. B}
  {\bfseries 804} (2020) 135377},
  \href{http://arxiv.org/abs/1910.09110}{{\ttfamily arXiv:1910.09110
  [nucl-ex]}}.

\bibitem{ALICE:2019rmo}
{\bfseries ALICE} Collaboration, S.~Acharya {\em et~al.}, ``{Production of
  muons from heavy-flavour hadron decays in pp collisions at $ \sqrt{s} $ =
  5.02 TeV}'', \href{http://dx.doi.org/10.1007/JHEP09(2019)008}{{\em JHEP}
  {\bfseries 09} (2019) 008}, \href{http://arxiv.org/abs/1905.07207}{{\ttfamily
  arXiv:1905.07207 [nucl-ex]}}.

\bibitem{ALICE:2021psx}
{\bfseries ALICE} Collaboration, S.~Acharya {\em et~al.}, ``{Measurement of the
  production cross section of prompt $ {\Xi}_{\mathrm{c}}^0 $ baryons at
  midrapidity in pp collisions at $ \sqrt{s} $ = 5.02 TeV}'',
  \href{http://dx.doi.org/10.1007/JHEP10(2021)159}{{\em JHEP} {\bfseries 10}
  (2021) 159}, \href{http://arxiv.org/abs/2105.05616}{{\ttfamily
  arXiv:2105.05616 [nucl-ex]}}.

\bibitem{ALICE:2021bli}
{\bfseries ALICE} Collaboration, S.~Acharya {\em et~al.}, ``{Measurement of the
  Cross Sections of $\Xi^0_{c}$ and $\Xi^+_{c}$ Baryons and of the
  Branching-Fraction Ratio BR($\Xi^0_{c} \rightarrow \Xi^-{e}^+\nu_{
  e}$)/BR($\Xi^0_{c} \rightarrow \Xi^-\pi^+$) in pp collisions at 13 TeV}'',
  \href{http://dx.doi.org/10.1103/PhysRevLett.127.272001}{{\em Phys. Rev.
  Lett.} {\bfseries 127} (2021) 272001},
  \href{http://arxiv.org/abs/2105.05187}{{\ttfamily arXiv:2105.05187
  [nucl-ex]}}.

\bibitem{ALICE:2020wfu}
{\bfseries ALICE} Collaboration, S.~Acharya {\em et~al.}, ``{$\Lambda^+_c$
  Production and Baryon-to-Meson Ratios in pp and p--Pb Collisions at $\sqrt
  {s_{\rm NN}}$=5.02\,\,TeV at the LHC}'',
  \href{http://dx.doi.org/10.1103/PhysRevLett.127.202301}{{\em Phys. Rev.
  Lett.} {\bfseries 127} (2021) 202301},
  \href{http://arxiv.org/abs/2011.06078}{{\ttfamily arXiv:2011.06078
  [nucl-ex]}}.

\bibitem{ALICE:2020wla}
{\bfseries ALICE} Collaboration, S.~Acharya {\em et~al.}, ``{$\Lambda^+_c$
  production in pp and in p--Pb collisions at $\sqrt {s_{\rm NN}}$=5.02 TeV}'',
  \href{http://dx.doi.org/10.1103/PhysRevC.104.054905}{{\em Phys. Rev. C}
  {\bfseries 104} (2021) 054905},
  \href{http://arxiv.org/abs/2011.06079}{{\ttfamily arXiv:2011.06079
  [nucl-ex]}}.

\bibitem{ALICE:2021rzj}
{\bfseries ALICE} Collaboration, S.~Acharya {\em et~al.}, ``{Measurement of
  Prompt D$^{0}$, $\Lambda_{c}^{+}$, and $\Sigma_{c}^{0,++}$(2455) Production
  in Proton\textendash{}Proton Collisions at $\sqrt s$ = 13\,\,TeV}'',
  \href{http://dx.doi.org/10.1103/PhysRevLett.128.012001}{{\em Phys. Rev.
  Lett.} {\bfseries 128} (2022) 012001},
  \href{http://arxiv.org/abs/2106.08278}{{\ttfamily arXiv:2106.08278
  [hep-ex]}}.

\bibitem{ALICE:2022cop}
{\bfseries ALICE} Collaboration, ``{First measurement of $\rm \Omega_c^0$
  production in pp collisions at $\sqrt{s}=13$ TeV}'',
  \href{http://dx.doi.org/10.1016/j.physletb.2022.137625}{{\em Phys. Lett. B}
  {\bfseries 846} (2023) 137625},
  \href{http://arxiv.org/abs/2205.13993}{{\ttfamily arXiv:2205.13993
  [nucl-ex]}}.

\bibitem{ATLAS:2012sfc}
{\bfseries ATLAS} Collaboration, G.~Aad {\em et~al.}, ``{Measurement of the
  b-hadron production cross section using decays to $D^{*}\mu^-X$ final states
  in pp collisions at sqrt(s) = 7 TeV with the ATLAS detector}'',
  \href{http://dx.doi.org/10.1016/j.nuclphysb.2012.07.009}{{\em Nucl. Phys. B}
  {\bfseries 864} (2012) 341--381},
  \href{http://arxiv.org/abs/1206.3122}{{\ttfamily arXiv:1206.3122 [hep-ex]}}.

\bibitem{ATLAS:2015igt}
{\bfseries ATLAS} Collaboration, G.~Aad {\em et~al.}, ``{Measurement of
  $D^{*\pm}$, $D^\pm$ and $D_s^\pm$ meson production cross sections in $pp$
  collisions at $\sqrt{s}=7$ TeV with the ATLAS detector}'',
  \href{http://dx.doi.org/10.1016/j.nuclphysb.2016.04.032}{{\em Nucl. Phys. B}
  {\bfseries 907} (2016) 717--763},
  \href{http://arxiv.org/abs/1512.02913}{{\ttfamily arXiv:1512.02913
  [hep-ex]}}.

\bibitem{ATLAS:2013cia}
{\bfseries ATLAS} Collaboration, G.~Aad {\em et~al.}, ``{Measurement of the
  differential cross-section of $B^{+}$ meson production in pp collisions at
  $\sqrt{s}$ = 7 TeV at ATLAS}'',
  \href{http://dx.doi.org/10.1007/JHEP10(2013)042}{{\em JHEP} {\bfseries 10}
  (2013) 042}, \href{http://arxiv.org/abs/1307.0126}{{\ttfamily arXiv:1307.0126
  [hep-ex]}}.

\bibitem{ATLAS:2015esn}
{\bfseries ATLAS} Collaboration, G.~Aad {\em et~al.}, ``{Determination of the
  ratio of $b$-quark fragmentation fractions $f_s/f_d$ in $pp$ collisions at
  $\sqrt{s}=7$ TeV with the ATLAS detector}'',
  \href{http://dx.doi.org/10.1103/PhysRevLett.115.262001}{{\em Phys. Rev.
  Lett.} {\bfseries 115} (2015) 262001},
  \href{http://arxiv.org/abs/1507.08925}{{\ttfamily arXiv:1507.08925
  [hep-ex]}}.

\bibitem{ATLAS:2019jpi}
{\bfseries ATLAS} Collaboration, M.~Aaboud {\em et~al.}, ``{Measurement of the
  relative $B^{\pm}_{c}/B^{\pm}$ production cross section with the ATLAS
  detector at $\sqrt{s}=8$ TeV}'',
  \href{http://dx.doi.org/10.1103/PhysRevD.104.012010}{{\em Phys. Rev. D}
  {\bfseries 104} (2021) 012010},
  \href{http://arxiv.org/abs/1912.02672}{{\ttfamily arXiv:1912.02672
  [hep-ex]}}.

\bibitem{CMS:2012xsp}
{\bfseries CMS} Collaboration, S.~Chatrchyan {\em et~al.}, ``{Measurement of
  the cross section for production of $b b^-$ bar $X$, decaying to muons in
  $pp$ collisions at $\sqrt{s}=7$ TeV}'',
  \href{http://dx.doi.org/10.1007/JHEP06(2012)110}{{\em JHEP} {\bfseries 06}
  (2012) 110}, \href{http://arxiv.org/abs/1203.3458}{{\ttfamily arXiv:1203.3458
  [hep-ex]}}.

\bibitem{CMS:2017qjw}
{\bfseries CMS} Collaboration, A.~M. Sirunyan {\em et~al.}, ``{Nuclear
  modification factor of D$^0$ mesons in PbPb collisions at
  $\sqrt{s_\mathrm{NN}} = 5.02$ TeV}'',
  \href{http://dx.doi.org/10.1016/j.physletb.2018.05.074}{{\em Phys. Lett. B}
  {\bfseries 782} (2018) 474--496},
  \href{http://arxiv.org/abs/1708.04962}{{\ttfamily arXiv:1708.04962
  [nucl-ex]}}.

\bibitem{CMS:2017uoy}
{\bfseries CMS} Collaboration, A.~M. Sirunyan {\em et~al.}, ``{Measurement of
  the ${B}^{\pm}$ Meson Nuclear Modification Factor in Pb-Pb Collisions at
  $\sqrt{{s}_{NN}}=5.02\text{ }\text{ }\mathrm{TeV}$}'',
  \href{http://dx.doi.org/10.1103/PhysRevLett.119.152301}{{\em Phys. Rev.
  Lett.} {\bfseries 119} (2017) 152301},
  \href{http://arxiv.org/abs/1705.04727}{{\ttfamily arXiv:1705.04727
  [hep-ex]}}.

\bibitem{CMS:2019uws}
{\bfseries CMS} Collaboration, A.~M. Sirunyan {\em et~al.}, ``{Production of
  $\Lambda_\mathrm{c}^+$ baryons in proton-proton and lead-lead collisions at
  $\sqrt{s_\mathrm{NN}}=$ 5.02 TeV}'',
  \href{http://dx.doi.org/10.1016/j.physletb.2020.135328}{{\em Phys. Lett. B}
  {\bfseries 803} (2020) 135328},
  \href{http://arxiv.org/abs/1906.03322}{{\ttfamily arXiv:1906.03322
  [hep-ex]}}.

\bibitem{CMS:2016plw}
{\bfseries CMS} Collaboration, V.~Khachatryan {\em et~al.}, ``{Measurement of
  the total and differential inclusive $B^+$ hadron cross sections in pp
  collisions at $\sqrt{s}$ = 13 TeV}'',
  \href{http://dx.doi.org/10.1016/j.physletb.2017.05.074}{{\em Phys. Lett. B}
  {\bfseries 771} (2017) 435--456},
  \href{http://arxiv.org/abs/1609.00873}{{\ttfamily arXiv:1609.00873
  [hep-ex]}}.

\bibitem{CMS:2018bwt}
{\bfseries CMS} Collaboration, A.~M. Sirunyan {\em et~al.}, ``{Studies of
  Beauty Suppression via Nonprompt $D^0$ Mesons in Pb-Pb Collisions at $Q^2 =
  4$ $\rm GeV^2$}'',
  \href{http://dx.doi.org/10.1103/PhysRevLett.123.022001}{{\em Phys. Rev.
  Lett.} {\bfseries 123} (2019) 022001},
  \href{http://arxiv.org/abs/1810.11102}{{\ttfamily arXiv:1810.11102
  [hep-ex]}}.

\bibitem{CMS:2014oqy}
{\bfseries CMS} Collaboration, V.~Khachatryan {\em et~al.}, ``{Measurement of
  the ratio of the production cross sections times branching fractions of
  $B_{c}^{\pm} \to J/\psi \pi^{\pm}$ and $B^{\pm} \to J/\psi K^{\pm}$ and
  $\mathcal{B}(B_{c}^{\pm} \to J/\psi
  \pi^{\pm}\pi^{\pm}\pi^{\mp})/\mathcal{B}(B_{c}^{\pm} \to J/\psi \pi^{\pm})$
  in pp collisions at $\sqrt{s} =$ 7 TeV}'',
  \href{http://dx.doi.org/10.1007/JHEP01(2015)063}{{\em JHEP} {\bfseries 01}
  (2015) 063}, \href{http://arxiv.org/abs/1410.5729}{{\ttfamily arXiv:1410.5729
  [hep-ex]}}.

\bibitem{LHCb:2013vjr}
{\bfseries LHCb} Collaboration, R.~Aaij {\em et~al.}, ``{Measurement of B meson
  production cross-sections in proton-proton collisions at $\sqrt{s}$ = 7
  TeV}'', \href{http://dx.doi.org/10.1007/JHEP08(2013)117}{{\em JHEP}
  {\bfseries 08} (2013) 117}, \href{http://arxiv.org/abs/1306.3663}{{\ttfamily
  arXiv:1306.3663 [hep-ex]}}.

\bibitem{LHCb:2016ikn}
{\bfseries LHCb} Collaboration, R.~Aaij {\em et~al.}, ``{Measurements of prompt
  charm production cross-sections in pp collisions at $ \sqrt{s}=5 $ TeV}'',
  \href{http://dx.doi.org/10.1007/JHEP06(2017)147}{{\em JHEP} {\bfseries 06}
  (2017) 147}, \href{http://arxiv.org/abs/1610.02230}{{\ttfamily
  arXiv:1610.02230 [hep-ex]}}.

\bibitem{LHCb:2015swx}
{\bfseries LHCb} Collaboration, R.~Aaij {\em et~al.}, ``{Measurements of prompt
  charm production cross-sections in $pp$ collisions at $ \sqrt{s}=13 $ TeV}'',
  \href{http://dx.doi.org/10.1007/JHEP03(2016)159}{{\em JHEP} {\bfseries 03}
  (2016) 159}, \href{http://arxiv.org/abs/1510.01707}{{\ttfamily
  arXiv:1510.01707 [hep-ex]}}. [Erratum: JHEP 09, 013 (2016), Erratum: JHEP 05,
  074 (2017)].

\bibitem{LHCb:2014mvo}
{\bfseries LHCb} Collaboration, R.~Aaij {\em et~al.}, ``{Measurement of $B_c^+$
  production in proton-proton collisions at $\sqrt{s}=8$ TeV}'',
  \href{http://dx.doi.org/10.1103/PhysRevLett.114.132001}{{\em Phys. Rev.
  Lett.} {\bfseries 114} (2015) 132001},
  \href{http://arxiv.org/abs/1411.2943}{{\ttfamily arXiv:1411.2943 [hep-ex]}}.

\bibitem{LHCb:2015qvk}
{\bfseries LHCb} Collaboration, R.~Aaij {\em et~al.}, ``{Study of the
  production of $\Lambda_b^0$ and $\overline{B}^0$ hadrons in $pp$ collisions
  and first measurement of the $\Lambda_b^0\rightarrow J/\psi pK^-$ branching
  fraction}'', \href{http://dx.doi.org/10.1088/1674-1137/40/1/011001}{{\em
  Chin. Phys. C} {\bfseries 40} (2016) 011001},
  \href{http://arxiv.org/abs/1509.00292}{{\ttfamily arXiv:1509.00292
  [hep-ex]}}.

\bibitem{LHCb:2016qpe}
{\bfseries LHCb} Collaboration, R.~Aaij {\em et~al.}, ``{Measurement of the
  $b$-quark production cross-section in 7 and 13 TeV $pp$ collisions}'',
  \href{http://dx.doi.org/10.1103/PhysRevLett.118.052002}{{\em Phys. Rev.
  Lett.} {\bfseries 118} (2017) 052002},
  \href{http://arxiv.org/abs/1612.05140}{{\ttfamily arXiv:1612.05140
  [hep-ex]}}. [Erratum: \textit{Phys. Rev. Lett.} \textbf{119}, (2017) 169901].

\bibitem{LHCb:2017vec}
{\bfseries LHCb} Collaboration, R.~Aaij {\em et~al.}, ``{Measurement of the
  $B^{\pm}$ production cross-section in pp collisions at $\sqrt{s} =$ 7 and 13
  TeV}'', \href{http://dx.doi.org/10.1007/JHEP12(2017)026}{{\em JHEP}
  {\bfseries 12} (2017) 026}, \href{http://arxiv.org/abs/1710.04921}{{\ttfamily
  arXiv:1710.04921 [hep-ex]}}.

\bibitem{LHCb:2019sxa}
{\bfseries LHCb} Collaboration, R.~Aaij {\em et~al.}, ``{Measurement of the
  mass and production rate of $\Xi_b^-$ baryons}'',
  \href{http://dx.doi.org/10.1103/PhysRevD.99.052006}{{\em Phys. Rev. D}
  {\bfseries 99} (2019) 052006},
  \href{http://arxiv.org/abs/1901.07075}{{\ttfamily arXiv:1901.07075
  [hep-ex]}}.

\bibitem{LHCb:2019fns}
{\bfseries LHCb} Collaboration, R.~Aaij {\em et~al.}, ``{Measurement of $b$
  hadron fractions in 13 TeV $pp$ collisions}'',
  \href{http://dx.doi.org/10.1103/PhysRevD.100.031102}{{\em Phys. Rev. D}
  {\bfseries 100} (2019) 031102},
  \href{http://arxiv.org/abs/1902.06794}{{\ttfamily arXiv:1902.06794
  [hep-ex]}}.

\bibitem{LHCb:2019qed}
{\bfseries LHCb} Collaboration, R.~Aaij {\em et~al.}, ``{Measurement of
  $\Xi_\mathrm{cc}^{++}$ production in $pp$ collisions at $\sqrt{s}=13$ TeV}'',
  \href{http://dx.doi.org/10.1088/1674-1137/44/2/022001}{{\em Chin. Phys. C}
  {\bfseries 44} (2020) 022001},
  \href{http://arxiv.org/abs/1910.11316}{{\ttfamily arXiv:1910.11316
  [hep-ex]}}.

\bibitem{STAR:2012nbd}
{\bfseries STAR} Collaboration, L.~Adamczyk {\em et~al.}, ``{Measurements of
  $D^{0}$ and $D^{*}$ Production in $p+p$ Collisions at $\sqrt{s} = 200$
  GeV}'', \href{http://dx.doi.org/10.1103/PhysRevD.86.072013}{{\em Phys. Rev.
  D} {\bfseries 86} (2012) 072013},
  \href{http://arxiv.org/abs/1204.4244}{{\ttfamily arXiv:1204.4244 [nucl-ex]}}.

\bibitem{Maciula:2013wg}
R.~Maciula and A.~Szczurek, ``{Open charm production at the LHC -
  $k_{t}$-factorization approach}'',
  \href{http://dx.doi.org/10.1103/PhysRevD.87.094022}{{\em Phys. Rev. D}
  {\bfseries 87} (2013) 094022},
  \href{http://arxiv.org/abs/1301.3033}{{\ttfamily arXiv:1301.3033 [hep-ph]}}.

\bibitem{Maciula:2018iuh}
R.~Maciu\l{}a and A.~Szczurek, ``{Production of $\Lambda_c$ baryons at the LHC
  within the $k_T$-factorization approach and independent parton fragmentation
  picture}'', \href{http://dx.doi.org/10.1103/PhysRevD.98.014016}{{\em Phys.
  Rev. D} {\bfseries 98} (2018) 014016},
  \href{http://arxiv.org/abs/1803.05807}{{\ttfamily arXiv:1803.05807
  [hep-ph]}}.

\bibitem{Guiot:2018kfy}
B.~Guiot, ``{Heavy-quark production with $k_t$-factorization: The importance of
  the sea-quark distribution}'',
  \href{http://dx.doi.org/10.1103/PhysRevD.99.074006}{{\em Phys. Rev. D}
  {\bfseries 99} (2019) 074006},
  \href{http://arxiv.org/abs/1812.02156}{{\ttfamily arXiv:1812.02156
  [hep-ph]}}.

\bibitem{Cacciari:1998it}
M.~Cacciari, M.~Greco, and P.~Nason, ``{The $p_{\rm T}$ spectrum in heavy
  flavor hadroproduction}'',
  \href{http://dx.doi.org/10.1088/1126-6708/1998/05/007}{{\em JHEP} {\bfseries
  05} (1998) 007}, \href{http://arxiv.org/abs/hep-ph/9803400}{{\ttfamily
  arXiv:hep-ph/9803400}}.

\bibitem{Cacciari:2001td}
M.~Cacciari, S.~Frixione, and P.~Nason, ``{The $p_{\rm T}$ spectrum in heavy
  flavor photoproduction}'',
  \href{http://dx.doi.org/10.1088/1126-6708/2001/03/006}{{\em JHEP} {\bfseries
  03} (2001) 006}, \href{http://arxiv.org/abs/hep-ph/0102134}{{\ttfamily
  arXiv:hep-ph/0102134}}.

\bibitem{Cacciari:2012ny}
M.~Cacciari {\em et~al.}, ``{Theoretical predictions for charm and bottom
  production at the LHC}'',
  \href{http://dx.doi.org/10.1007/JHEP10(2012)137}{{\em JHEP} {\bfseries 10}
  (2012) 137}, \href{http://arxiv.org/abs/1205.6344}{{\ttfamily arXiv:1205.6344
  [hep-ph]}}.

\bibitem{Kniehl:2004fy}
B.~Kniehl, G.~Kramer, I.~Schienbein, and H.~Spiesberger, ``{Inclusive
  ${D}^{*\ifmmode\pm\else\textpm\fi{}}$ production in $p\overline{p}$
  collisions with massive charm quarks}'',
  \href{http://dx.doi.org/10.1103/PhysRevD.71.014018}{{\em Phys. Rev. D}
  {\bfseries 71} (2005) 014018},
  \href{http://arxiv.org/abs/hep-ph/0410289}{{\ttfamily arXiv:hep-ph/0410289}}.

\bibitem{Kniehl:2012ti}
B.~Kniehl, G.~Kramer, I.~Schienbein, and H.~Spiesberger, ``{Inclusive
  Charmed-Meson Production at the CERN LHC}'',
  \href{http://dx.doi.org/10.1140/epjc/s10052-012-2082-2}{{\em Eur. Phys. J. C}
  {\bfseries 72} (2012) 2082}, \href{http://arxiv.org/abs/1202.0439}{{\ttfamily
  arXiv:1202.0439 [hep-ph]}}.

\bibitem{Benzke:2017yjn}
M.~Benzke {\em et~al.}, ``{Prompt neutrinos from atmospheric charm in the
  general-mass variable-flavor-number scheme}'',
  \href{http://dx.doi.org/10.1007/JHEP12(2017)021}{{\em JHEP} {\bfseries 12}
  (2017) 021}, \href{http://arxiv.org/abs/1705.10386}{{\ttfamily
  arXiv:1705.10386 [hep-ph]}}.

\bibitem{Kramer:2017gct}
G.~Kramer and H.~Spiesberger, ``{Study of heavy meson production in
  p\textendash{}Pb collisions at $\sqrt{S}$= 5.02 TeV in the general-mass
  variable-flavour-number scheme}'',
  \href{http://dx.doi.org/10.1016/j.nuclphysb.2017.10.016}{{\em Nucl. Phys. B}
  {\bfseries 925} (2017) 415--430},
  \href{http://arxiv.org/abs/1703.04754}{{\ttfamily arXiv:1703.04754
  [hep-ph]}}.

\bibitem{Helenius:2018uul}
I.~Helenius and H.~Paukkunen, ``{Revisiting the D-meson hadroproduction in
  general-mass variable flavour number scheme}'',
  \href{http://dx.doi.org/10.1007/JHEP05(2018)196}{{\em JHEP} {\bfseries 05}
  (2018) 196}, \href{http://arxiv.org/abs/1804.03557}{{\ttfamily
  arXiv:1804.03557 [hep-ph]}}.

\bibitem{Bolzoni:2013vya}
P.~Bolzoni and G.~Kramer, ``{Inclusive charmed-meson production from bottom
  hadron decays at the LHC}'',
  \href{http://dx.doi.org/10.1088/0954-3899/41/7/075006}{{\em J. Phys. G}
  {\bfseries 41} (2014) 075006},
  \href{http://arxiv.org/abs/1310.2924}{{\ttfamily arXiv:1310.2924 [hep-ph]}}.

\bibitem{Braaten:1994bz}
E.~Braaten, K.-m. Cheung, S.~Fleming, and T.~C. Yuan, ``{Perturbative QCD
  fragmentation functions as a model for heavy quark fragmentation}'',
  \href{http://dx.doi.org/10.1103/PhysRevD.51.4819}{{\em Phys. Rev. D}
  {\bfseries 51} (1995) 4819--4829},
  \href{http://arxiv.org/abs/hep-ph/9409316}{{\ttfamily arXiv:hep-ph/9409316}}.

\bibitem{ALICE:2021dhb}
{\bfseries ALICE} Collaboration, S.~Acharya {\em et~al.}, ``{Charm-quark
  fragmentation fractions and production cross section at midrapidity in pp
  collisions at the LHC}'',
  \href{http://dx.doi.org/10.1103/PhysRevD.105.L011103}{{\em Phys. Rev. D}
  {\bfseries 105} (2022) L011103},
  \href{http://arxiv.org/abs/2105.06335}{{\ttfamily arXiv:2105.06335
  [nucl-ex]}}.

\bibitem{Sjostrand:2014zea}
T.~Sj\"ostrand {\em et~al.}, ``{An introduction to PYTHIA 8.2}'',
  \href{http://dx.doi.org/10.1016/j.cpc.2015.01.024}{{\em Comput. Phys.
  Commun.} {\bfseries 191} (2015) 159--177},
  \href{http://arxiv.org/abs/1410.3012}{{\ttfamily arXiv:1410.3012 [hep-ph]}}.

\bibitem{Skands:2014pea}
P.~Skands, S.~Carrazza, and J.~Rojo, ``{Tuning PYTHIA 8.1: the Monash 2013
  Tune}'', \href{http://dx.doi.org/10.1140/epjc/s10052-014-3024-y}{{\em Eur.
  Phys. J. C} {\bfseries 74} (2014) 3024},
  \href{http://arxiv.org/abs/1404.5630}{{\ttfamily arXiv:1404.5630 [hep-ph]}}.

\bibitem{Bellm:2015jjp}
J.~Bellm {\em et~al.}, ``{Herwig 7.0/Herwig++ 3.0 release note}'',
  \href{http://dx.doi.org/10.1140/epjc/s10052-016-4018-8}{{\em Eur. Phys. J. C}
  {\bfseries 76} (2016) 196}, \href{http://arxiv.org/abs/1512.01178}{{\ttfamily
  arXiv:1512.01178 [hep-ph]}}.

\bibitem{Christiansen:2015yqa}
J.~R. Christiansen and P.~Z. Skands, ``{String Formation Beyond Leading
  Colour}'', \href{http://dx.doi.org/10.1007/JHEP08(2015)003}{{\em JHEP}
  {\bfseries 08} (2015) 003}, \href{http://arxiv.org/abs/1505.01681}{{\ttfamily
  arXiv:1505.01681 [hep-ph]}}.

\bibitem{ALICE:2019avo}
{\bfseries ALICE} Collaboration, S.~Acharya {\em et~al.}, ``{Multiplicity
  dependence of (multi-)strange hadron production in proton-proton collisions
  at $\sqrt{s}$ = 13 TeV}'',
  \href{http://dx.doi.org/10.1140/epjc/s10052-020-7673-8}{{\em Eur. Phys. J. C}
  {\bfseries 80} (2020) 167}, \href{http://arxiv.org/abs/1908.01861}{{\ttfamily
  arXiv:1908.01861 [nucl-ex]}}.

\bibitem{Bierlich:2014xba}
C.~Bierlich, G.~Gustafson, L.~L\"onnblad, and A.~Tarasov, ``{Effects of
  Overlapping Strings in pp Collisions}'',
  \href{http://dx.doi.org/10.1007/JHEP03(2015)148}{{\em JHEP} {\bfseries 03}
  (2015) 148}, \href{http://arxiv.org/abs/1412.6259}{{\ttfamily arXiv:1412.6259
  [hep-ph]}}.

\bibitem{Sjostrand:1987su}
T.~Sjostrand and M.~van Zijl, ``{A Multiple Interaction Model for the Event
  Structure in Hadron Collisions}'',
  \href{http://dx.doi.org/10.1103/PhysRevD.36.2019}{{\em Phys. Rev. D}
  {\bfseries 36} (1987) 2019}.

\bibitem{Bartalini:2009quk}
P.~Bartalini and L.~Fano, eds.,
  \href{http://dx.doi.org/10.3204/PUBDB-2017-128830}{{\em {Proceedings, 1st
  International Workshop on Multiple Partonic Interactions at the LHC (MPI08)}:
  {Perugia, Italy, October 27-31, 2008}}}.
\newblock DESY, Hamburg, 2009.
\newblock \href{http://arxiv.org/abs/1003.4220}{{\ttfamily arXiv:1003.4220
  [hep-ex]}}.

\bibitem{Schmidt:2020fgn}
I.~Schmidt and M.~Siddikov, ``{Production mechanisms of open-heavy flavor
  mesons}'', \href{http://dx.doi.org/10.1103/PhysRevD.101.094020}{{\em Phys.
  Rev. D} {\bfseries 101} (2020) 094020},
  \href{http://arxiv.org/abs/2003.13768}{{\ttfamily arXiv:2003.13768
  [hep-ph]}}.

\bibitem{ALICE:2015ikl}
{\bfseries ALICE} Collaboration, J.~Adam {\em et~al.}, ``{Measurement of charm
  and beauty production at central rapidity versus charged-particle
  multiplicity in proton-proton collisions at $ \sqrt{s}=7 $ TeV}'',
  \href{http://dx.doi.org/10.1007/JHEP09(2015)148}{{\em JHEP} {\bfseries 09}
  (2015) 148}, \href{http://arxiv.org/abs/1505.00664}{{\ttfamily
  arXiv:1505.00664 [nucl-ex]}}.

\bibitem{ALICE:2020msa}
{\bfseries ALICE} Collaboration, S.~Acharya {\em et~al.}, ``{Multiplicity
  dependence of J/$\psi$ production at midrapidity in pp collisions at
  $\sqrt{s}$ = 13 TeV}'',
  \href{http://dx.doi.org/10.1016/j.physletb.2020.135758}{{\em Phys. Lett. B}
  {\bfseries 810} (2020) 135758},
  \href{http://arxiv.org/abs/2005.11123}{{\ttfamily arXiv:2005.11123
  [nucl-ex]}}.

\bibitem{ALICE:2021zkd}
{\bfseries ALICE} Collaboration, S.~Acharya {\em et~al.}, ``{Forward rapidity
  J/\ensuremath{\psi} production as a function of charged-particle multiplicity
  in pp collisions at $ \sqrt{s} $ = 5.02 and 13 TeV}'',
  \href{http://dx.doi.org/10.1007/JHEP06(2022)015}{{\em JHEP} {\bfseries 06}
  (2022) 015}, \href{http://arxiv.org/abs/2112.09433}{{\ttfamily
  arXiv:2112.09433 [nucl-ex]}}.

\bibitem{Werner:2013tya}
K.~Werner, B.~Guiot, I.~Karpenko, and T.~Pierog, ``{Analysing radial flow
  features in p-Pb and p-p collisions at several TeV by studying identified
  particle production in EPOS3}'',
  \href{http://dx.doi.org/10.1103/PhysRevC.89.064903}{{\em Phys. Rev. C}
  {\bfseries 89} (2014) 064903},
  \href{http://arxiv.org/abs/1312.1233}{{\ttfamily arXiv:1312.1233 [nucl-th]}}.

\bibitem{Braun-Munzinger:2015hba}
P.~Braun-Munzinger, V.~Koch, T.~Sch\"afer, and J.~Stachel, ``{Properties of hot
  and dense matter from relativistic heavy ion collisions}'',
  \href{http://dx.doi.org/10.1016/j.physrep.2015.12.003}{{\em Phys. Rept.}
  {\bfseries 621} (2016) 76--126},
  \href{http://arxiv.org/abs/1510.00442}{{\ttfamily arXiv:1510.00442
  [nucl-th]}}.

\bibitem{CMS:2010ifv}
{\bfseries CMS} Collaboration, V.~Khachatryan {\em et~al.}, ``{Observation of
  Long-Range Near-Side Angular Correlations in Proton-Proton Collisions at the
  LHC}'', \href{http://dx.doi.org/10.1007/JHEP09(2010)091}{{\em JHEP}
  {\bfseries 09} (2010) 091}, \href{http://arxiv.org/abs/1009.4122}{{\ttfamily
  arXiv:1009.4122 [hep-ex]}}.

\bibitem{ATLAS:2015hzw}
{\bfseries ATLAS} Collaboration, G.~Aad {\em et~al.}, ``{Observation of
  Long-Range Elliptic Azimuthal Anisotropies in $\sqrt{s}=$13 and 2.76 TeV $pp$
  Collisions with the ATLAS Detector}'',
  \href{http://dx.doi.org/10.1103/PhysRevLett.116.172301}{{\em Phys. Rev.
  Lett.} {\bfseries 116} (2016) 172301},
  \href{http://arxiv.org/abs/1509.04776}{{\ttfamily arXiv:1509.04776
  [hep-ex]}}.

\bibitem{ALICE:2016fzo}
{\bfseries ALICE} Collaboration, J.~Adam {\em et~al.}, ``{Enhanced production
  of multi-strange hadrons in high-multiplicity proton-proton collisions}'',
  \href{http://dx.doi.org/10.1038/nphys4111}{{\em Nature Phys.} {\bfseries 13}
  (2017) 535--539}, \href{http://arxiv.org/abs/1606.07424}{{\ttfamily
  arXiv:1606.07424 [nucl-ex]}}.

\bibitem{Chen:2020drg}
Y.~Chen and M.~He, ``{Charged-particle multiplicity dependence of
  charm-baryon-to-meson ratio in high-energy proton-proton collisions}'',
  \href{http://dx.doi.org/10.1016/j.physletb.2021.136144}{{\em Phys. Lett. B}
  {\bfseries 815} (2021) 136144},
  \href{http://arxiv.org/abs/2011.14328}{{\ttfamily arXiv:2011.14328
  [hep-ph]}}.

\bibitem{Minissale:2020bif}
V.~Minissale, S.~Plumari, and V.~Greco, ``{Charm hadrons in pp collisions at
  LHC energy within a coalescence plus fragmentation approach}'',
  \href{http://dx.doi.org/10.1016/j.physletb.2021.136622}{{\em Phys. Lett. B}
  {\bfseries 821} (2021) 136622},
  \href{http://arxiv.org/abs/2012.12001}{{\ttfamily arXiv:2012.12001
  [hep-ph]}}.

\bibitem{Plumari:2017ntm}
S.~Plumari, V.~Minissale, S.~K. Das, G.~Coci, and V.~Greco, ``{Charmed Hadrons
  from Coalescence plus Fragmentation in relativistic nucleus-nucleus
  collisions at RHIC and LHC}'',
  \href{http://dx.doi.org/10.1140/epjc/s10052-018-5828-7}{{\em Eur. Phys. J. C}
  {\bfseries 78} (2018) 348}, \href{http://arxiv.org/abs/1712.00730}{{\ttfamily
  arXiv:1712.00730 [hep-ph]}}.

\bibitem{ALICE:2021npz}
{\bfseries ALICE} Collaboration, S.~Acharya {\em et~al.}, ``{Observation of a
  multiplicity dependence in the $p_{\rm T}$-differential charm baryon-to-meson
  ratios in proton-proton collisions at $\sqrt{s} = 13$ TeV}'',
  \href{http://dx.doi.org/10.1016/j.physletb.2022.137065}{{\em Phys. Lett. B}
  {\bfseries 829} (2022) 137065},
  \href{http://arxiv.org/abs/2111.11948}{{\ttfamily arXiv:2111.11948
  [nucl-ex]}}.

\bibitem{ALICE:2021bib}
{\bfseries ALICE} Collaboration, S.~Acharya {\em et~al.}, ``{Constraining
  hadronization mechanisms with $\rm \Lambda_{\rm c}^{+}$/D$^0$ production
  ratios in Pb-Pb collisions at $\sqrt{s_{\rm NN}} = 5.02$ TeV}'',
  \href{http://dx.doi.org/10.1016/j.physletb.2023.137796}{{\em Phys. Lett. B}
  {\bfseries 839} (2023) 137796},
  \href{http://arxiv.org/abs/2112.08156}{{\ttfamily arXiv:2112.08156
  [nucl-ex]}}.

\bibitem{LHCb:2022syj}
{\bfseries LHCb} Collaboration, ``{Evidence for modification of $b$ quark
  hadronization in high-multiplicity $pp$ collisions at $\sqrt{s} = 13$ TeV}'',
  \href{http://dx.doi.org/10.1103/PhysRevLett.131.061901}{{\em Phys. Rev.
  Lett.} {\bfseries 131} (2023) 061901},
  \href{http://arxiv.org/abs/2204.13042}{{\ttfamily arXiv:2204.13042
  [hep-ex]}}.

\bibitem{LHCb:2020sey}
{\bfseries LHCb} Collaboration, R.~Aaij {\em et~al.}, ``{Observation of
  Multiplicity Dependent Prompt $\chi_{c1}(3872)$ and $\psi(2S)$ Production in
  $pp$ Collisions}'',
  \href{http://dx.doi.org/10.1103/PhysRevLett.126.092001}{{\em Phys. Rev.
  Lett.} {\bfseries 126} (2021) 092001},
  \href{http://arxiv.org/abs/2009.06619}{{\ttfamily arXiv:2009.06619
  [hep-ex]}}.

\bibitem{Esposito:2020ywk}
A.~Esposito, E.~G. Ferreiro, A.~Pilloni, A.~D. Polosa, and C.~A. Salgado,
  ``{The nature of X(3872) from high-multiplicity pp collisions}'',
  \href{http://dx.doi.org/10.1140/epjc/s10052-021-09425-w}{{\em Eur. Phys. J.
  C} {\bfseries 81} (2021) 669},
  \href{http://arxiv.org/abs/2006.15044}{{\ttfamily arXiv:2006.15044
  [hep-ph]}}.

\bibitem{Braaten:2020iqw}
E.~Braaten, L.-P. He, K.~Ingles, and J.~Jiang, ``{Production of $X(3872)$ at
  High Multiplicity}'',
  \href{http://dx.doi.org/10.1103/PhysRevD.103.L071901}{{\em Phys. Rev. D}
  {\bfseries 103} (2021) L071901},
  \href{http://arxiv.org/abs/2012.13499}{{\ttfamily arXiv:2012.13499
  [hep-ph]}}.

\bibitem{ALICE:2014sbx}
{\bfseries ALICE} Collaboration, B.~B. Abelev {\em et~al.}, ``{Performance of
  the ALICE Experiment at the CERN LHC}'',
  \href{http://dx.doi.org/10.1142/S0217751X14300440}{{\em Int. J. Mod. Phys. A}
  {\bfseries 29} (2014) 1430044},
  \href{http://arxiv.org/abs/1402.4476}{{\ttfamily arXiv:1402.4476 [nucl-ex]}}.

\bibitem{ALICE:2008ngc}
{\bfseries ALICE} Collaboration, K.~Aamodt {\em et~al.}, ``{The ALICE
  experiment at the CERN LHC}'',
  \href{http://dx.doi.org/10.1088/1748-0221/3/08/S08002}{{\em JINST} {\bfseries
  3} (2008) S08002}.

\bibitem{ALICE:2020swj}
{\bfseries ALICE} Collaboration, S.~Acharya {\em et~al.}, ``{Pseudorapidity
  distributions of charged particles as a function of mid- and forward rapidity
  multiplicities in pp collisions at $\sqrt{s}$~=~5.02, 7 and 13 TeV}'',
  \href{http://dx.doi.org/10.1140/epjc/s10052-021-09349-5}{{\em Eur. Phys. J.
  C} {\bfseries 81} (2021) 630},
  \href{http://arxiv.org/abs/2009.09434}{{\ttfamily arXiv:2009.09434
  [nucl-ex]}}.

\bibitem{ALICE:2021leo}
{\bfseries ALICE} Collaboration, S.~Acharya {\em et~al.}, ``{ALICE
  2016-2017-2018 luminosity determination for pp collisions at $\sqrt{s}=13$
  TeV}'',. \url{https://cds.cern.ch/record/2776672/}.

\bibitem{Sjostrand:2006za}
T.~Sjostrand, S.~Mrenna, and P.~Z. Skands, ``{PYTHIA 6.4 Physics and Manual}'',
  \href{http://dx.doi.org/10.1088/1126-6708/2006/05/026}{{\em JHEP} {\bfseries
  05} (2006) 026}, \href{http://arxiv.org/abs/hep-ph/0603175}{{\ttfamily
  arXiv:hep-ph/0603175}}.

\bibitem{Brun:1994aa}
R.~Brun {\em et~al.}, \href{http://dx.doi.org/10.17181/CERN.MUHF.DMJ1}{{\em
  {GEANT: Detector Description and Simulation Tool; Oct 1994}}}.
\newblock CERN Program Library. CERN, Geneva, 1993.
\newblock \url{http://cds.cern.ch/record/1082634}.
\newblock Long Writeup W5013.

\bibitem{ParticleDataGroup:2022pth}
{\bfseries Particle Data Group} Collaboration, R.~L. Workman {\em et~al.},
  ``{Review of Particle Physics}'',
  \href{http://dx.doi.org/10.1093/ptep/ptac097}{{\em PTEP} {\bfseries 2022}
  (2022) 083C01}.

\bibitem{Chen:2016XST}
T.~Chen and C.~Guestrin, ``Xgboost: A scalable tree boosting system'',
  \href{http://dx.doi.org/10.1145/2939672.2939785}{{\em Proceedings of the 22nd
  ACM SIGKDD International Conference on Knowledge Discovery and Data Mining}
  (2016) 785–794}, \href{http://arxiv.org/abs/1603.02754}{{\ttfamily
  arXiv:1603.02754 [cs.LG]}}.

\bibitem{barioglio_luca_2021_5070132}
L.~Barioglio, F.~Catalano, M.~Concas, P.~Fecchio, F.~Grosa, F.~Mazzaschi, and
  M.~Puccio, ``hipe4ml/hipe4ml'', July, 2021.
\newblock \url{https://doi.org/10.5281/zenodo.5070132}.

\bibitem{Bierlich:2022pfr}
C.~Bierlich {\em et~al.}, ``{A comprehensive guide to the physics and usage of
  PYTHIA 8.3}'', \href{http://arxiv.org/abs/2203.11601}{{\ttfamily
  arXiv:2203.11601 [hep-ph]}}.

\bibitem{Werner:2023zvo}
K.~Werner, ``{On a deep connection between factorization and saturation: new
  insight into modeling high-energy proton-proton and nucleus-nucleus
  scattering in the EPOS4 framework}'',
  \href{http://arxiv.org/abs/2301.12517}{{\ttfamily arXiv:2301.12517
  [hep-ph]}}.

\bibitem{Kneesch:2007ey}
T.~Kneesch, B.~A. Kniehl, G.~Kramer, and I.~Schienbein, ``{Charmed-meson
  fragmentation functions with finite-mass corrections}'',
  \href{http://dx.doi.org/10.1016/j.nuclphysb.2008.02.015}{{\em Nucl. Phys. B}
  {\bfseries 799} (2008) 34--59},
  \href{http://arxiv.org/abs/0712.0481}{{\ttfamily arXiv:0712.0481 [hep-ph]}}.

\end{thebibliography}\endgroup

\newpage
\appendix

%
%

\section{The ALICE Collaboration}
\label{app:collab}
\begin{flushleft} 
\small

S.~Acharya\,\orcidlink{0000-0002-9213-5329}\,$^{\rm 125}$, 
D.~Adamov\'{a}\,\orcidlink{0000-0002-0504-7428}\,$^{\rm 86}$, 
A.~Adler$^{\rm 70}$, 
G.~Aglieri Rinella\,\orcidlink{0000-0002-9611-3696}\,$^{\rm 33}$, 
M.~Agnello\,\orcidlink{0000-0002-0760-5075}\,$^{\rm 30}$, 
N.~Agrawal\,\orcidlink{0000-0003-0348-9836}\,$^{\rm 51}$, 
Z.~Ahammed\,\orcidlink{0000-0001-5241-7412}\,$^{\rm 133}$, 
S.~Ahmad\,\orcidlink{0000-0003-0497-5705}\,$^{\rm 16}$, 
S.U.~Ahn\,\orcidlink{0000-0001-8847-489X}\,$^{\rm 71}$, 
I.~Ahuja\,\orcidlink{0000-0002-4417-1392}\,$^{\rm 38}$
% ...

\section*{Affiliation Notes}

$^{\rm I}$ Deceased\\
$^{\rm II}$ Also at: Max-Planck-Institut f\"{u}r Physik, Munich, Germany\\
$^{\rm III}$ Also at: Italian National Agency for New Technologies, Energy and Sustainable Economic Development (ENEA), Bologna, Italy\\
$^{\rm IV}$ Also at: Dipartimento DET del Politecnico di Torino, Turin, Italy\\
$^{\rm V}$ Also at: Department of Applied Physics, Aligarh Muslim University, Aligarh, India\\
$^{\rm VI}$ Also at: Institute of Theoretical Physics, University of Wroclaw, Poland\\
$^{\rm VII}$ Also at: An institution covered by a cooperation agreement with CERN\\

\section*{Collaboration Institutes}

$^{1}$ A.I. Alikhanyan National Science Laboratory (Yerevan Physics Institute) Foundation, Yerevan, Armenia\\
$^{2}$ AGH University of Krakow, Cracow, Poland\\
$^{3}$ Bogolyubov Institute for Theoretical Physics, National Academy of Sciences of Ukraine, Kiev, Ukraine\\
$^{4}$ Bose Institute, Department of Physics  and Centre for Astroparticle Physics and Space Science (CAPSS), Kolkata, India\\
$^{5}$ California Polytechnic State University, San Luis Obispo, California, United States\\
$^{6}$ Central China Normal University, Wuhan, China\\
$^{7}$ Centro de Aplicaciones Tecnol\'{o}gicas y Desarrollo Nuclear (CEADEN), Havana, Cuba\\
$^{8}$ Centro de Investigaci\'{o}n y de Estudios Avanzados (CINVESTAV), Mexico City and M\'{e}rida, Mexico\\
$^{9}$ Chicago State University, Chicago, Illinois, United States\\
$^{10}$ China Institute of Atomic Energy, Beijing, China\\
$^{11}$ China University of Geosciences, Wuhan, China\\
$^{12}$ Chungbuk National University, Cheongju, Republic of Korea\\
$^{13}$ Comenius University Bratislava, Faculty of Mathematics, Physics and Informatics, Bratislava, Slovak Republic\\
$^{14}$ COMSATS University Islamabad, Islamabad, Pakistan\\
$^{15}$ Creighton University, Omaha, Nebraska, United States\\
$^{16}$ Department of Physics, Aligarh Muslim University, Aligarh, India\\
$^{17}$ Department of Physics, Pusan National University, Pusan, Republic of Korea\\
$^{18}$ Department of Physics, Sejong University, Seoul, Republic of Korea\\
$^{19}$ Department of Physics, University of California, Berkeley, California, United States\\
$^{20}$ Department of Physics, University of Oslo, Oslo, Norway\\
$^{21}$ Department of Physics and Technology, University of Bergen, Bergen, Norway\\
$^{22}$ Dipartimento di Fisica, Universit\`{a} di Pavia, Pavia, Italy\\
$^{23}$ Dipartimento di Fisica dell'Universit\`{a} and Sezione INFN, Cagliari, Italy\\
$^{24}$ Dipartimento di Fisica dell'Universit\`{a} and Sezione INFN, Trieste, Italy\\
$^{25}$ Dipartimento di Fisica dell'Universit\`{a} and Sezione INFN, Turin, Italy\\
$^{26}$ Dipartimento di Fisica e Astronomia dell'Universit\`{a} and Sezione INFN, Bologna, Italy\\
$^{27}$ Dipartimento di Fisica e Astronomia dell'Universit\`{a} and Sezione INFN, Catania, Italy\\
$^{28}$ Dipartimento di Fisica e Astronomia dell'Universit\`{a} and Sezione INFN, Padova, Italy\\
$^{29}$ Dipartimento di Fisica `E.R.~Caianiello' dell'Universit\`{a} and Gruppo Collegato INFN, Salerno, Italy\\
$^{30}$ Dipartimento DISAT del Politecnico and Sezione INFN, Turin, Italy\\
$^{31}$ Dipartimento di Scienze MIFT, Universit\`{a} di Messina, Messina, Italy\\
$^{32}$ Dipartimento Interateneo di Fisica `M.~Merlin' and Sezione INFN, Bari, Italy\\
$^{33}$ European Organization for Nuclear Research (CERN), Geneva, Switzerland\\
$^{34}$ Faculty of Electrical Engineering, Mechanical Engineering and Naval Architecture, University of Split, Split, Croatia\\
$^{35}$ Faculty of Engineering and Science, Western Norway University of Applied Sciences, Bergen, Norway\\
$^{36}$ Faculty of Nuclear Sciences and Physical Engineering, Czech Technical University in Prague, Prague, Czech Republic\\
$^{37}$ Faculty of Physics, Sofia University, Sofia, Bulgaria\\
$^{38}$ Faculty of Science, P.J.~\v{S}af\'{a}rik University, Ko\v{s}ice, Slovak Republic\\
$^{39}$ Frankfurt Institute for Advanced Studies, Johann Wolfgang Goethe-Universit\"{a}t Frankfurt, Frankfurt, Germany\\
$^{40}$ Fudan University, Shanghai, China\\
$^{41}$ Gangneung-Wonju National University, Gangneung, Republic of Korea\\
$^{42}$ Gauhati University, Department of Physics, Guwahati, India\\
$^{43}$ Helmholtz-Institut f\"{u}r Strahlen- und Kernphysik, Rheinische Friedrich-Wilhelms-Universit\"{a}t Bonn, Bonn, Germany\\
$^{44}$ Helsinki Institute of Physics (HIP), Helsinki, Finland\\
$^{45}$ High Energy Physics Group,  Universidad Aut\'{o}noma de Puebla, Puebla, Mexico\\
$^{46}$ Horia Hulubei National Institute of Physics and Nuclear Engineering, Bucharest, Romania\\
$^{47}$ Indian Institute of Technology Bombay (IIT), Mumbai, India\\
$^{48}$ Indian Institute of Technology Indore, Indore, India\\
$^{49}$ INFN, Laboratori Nazionali di Frascati, Frascati, Italy\\
$^{50}$ INFN, Sezione di Bari, Bari, Italy\\
$^{51}$ INFN, Sezione di Bologna, Bologna, Italy\\
$^{52}$ INFN, Sezione di Cagliari, Cagliari, Italy\\
$^{53}$ INFN, Sezione di Catania, Catania, Italy\\
$^{54}$ INFN, Sezione di Padova, Padova, Italy\\
$^{55}$ INFN, Sezione di Pavia, Pavia, Italy\\
$^{56}$ INFN, Sezione di Torino, Turin, Italy\\
$^{57}$ INFN, Sezione di Trieste, Trieste, Italy\\
$^{58}$ Inha University, Incheon, Republic of Korea\\
$^{59}$ Institute for Gravitational and Subatomic Physics (GRASP), Utrecht University/Nikhef, Utrecht, Netherlands\\
$^{60}$ Institute of Experimental Physics, Slovak Academy of Sciences, Ko\v{s}ice, Slovak Republic\\
$^{61}$ Institute of Physics, Homi Bhabha National Institute, Bhubaneswar, India\\
$^{62}$ Institute of Physics of the Czech Academy of Sciences, Prague, Czech Republic\\
$^{63}$ Institute of Space Science (ISS), Bucharest, Romania\\
$^{64}$ Institut f\"{u}r Kernphysik, Johann Wolfgang Goethe-Universit\"{a}t Frankfurt, Frankfurt, Germany\\
$^{65}$ Instituto de Ciencias Nucleares, Universidad Nacional Aut\'{o}noma de M\'{e}xico, Mexico City, Mexico\\
$^{66}$ Instituto de F\'{i}sica, Universidade Federal do Rio Grande do Sul (UFRGS), Porto Alegre, Brazil\\
$^{67}$ Instituto de F\'{\i}sica, Universidad Nacional Aut\'{o}noma de M\'{e}xico, Mexico City, Mexico\\
$^{68}$ iThemba LABS, National Research Foundation, Somerset West, South Africa\\
$^{69}$ Jeonbuk National University, Jeonju, Republic of Korea\\
$^{70}$ Johann-Wolfgang-Goethe Universit\"{a}t Frankfurt Institut f\"{u}r Informatik, Fachbereich Informatik und Mathematik, Frankfurt, Germany\\
$^{71}$ Korea Institute of Science and Technology Information, Daejeon, Republic of Korea\\
$^{72}$ KTO Karatay University, Konya, Turkey\\
$^{73}$ Laboratoire de Physique Subatomique et de Cosmologie, Universit\'{e} Grenoble-Alpes, CNRS-IN2P3, Grenoble, France\\
$^{74}$ Lawrence Berkeley National Laboratory, Berkeley, California, United States\\
$^{75}$ Lund University Department of Physics, Division of Particle Physics, Lund, Sweden\\
$^{76}$ Nagasaki Institute of Applied Science, Nagasaki, Japan\\
$^{77}$ Nara Women{'}s University (NWU), Nara, Japan\\
$^{78}$ National and Kapodistrian University of Athens, School of Science, Department of Physics , Athens, Greece\\
$^{79}$ National Centre for Nuclear Research, Warsaw, Poland\\
$^{80}$ National Institute of Science Education and Research, Homi Bhabha National Institute, Jatni, India\\
$^{81}$ National Nuclear Research Center, Baku, Azerbaijan\\
$^{82}$ National Research and Innovation Agency - BRIN, Jakarta, Indonesia\\
$^{83}$ Niels Bohr Institute, University of Copenhagen, Copenhagen, Denmark\\
$^{84}$ Nikhef, National institute for subatomic physics, Amsterdam, Netherlands\\
$^{85}$ Nuclear Physics Group, STFC Daresbury Laboratory, Daresbury, United Kingdom\\
$^{86}$ Nuclear Physics Institute of the Czech Academy of Sciences, Husinec-\v{R}e\v{z}, Czech Republic\\
$^{87}$ Oak Ridge National Laboratory, Oak Ridge, Tennessee, United States\\
$^{88}$ Ohio State University, Columbus, Ohio, United States\\
$^{89}$ Physics department, Faculty of science, University of Zagreb, Zagreb, Croatia\\
$^{90}$ Physics Department, Panjab University, Chandigarh, India\\
$^{91}$ Physics Department, University of Jammu, Jammu, India\\
$^{92}$ Physics Program and International Institute for Sustainability with Knotted Chiral Meta Matter (SKCM2), Hiroshima University, Hiroshima, Japan\\
$^{93}$ Physikalisches Institut, Eberhard-Karls-Universit\"{a}t T\"{u}bingen, T\"{u}bingen, Germany\\
$^{94}$ Physikalisches Institut, Ruprecht-Karls-Universit\"{a}t Heidelberg, Heidelberg, Germany\\
$^{95}$ Physik Department, Technische Universit\"{a}t M\"{u}nchen, Munich, Germany\\
$^{96}$ Politecnico di Bari and Sezione INFN, Bari, Italy\\
$^{97}$ Research Division and ExtreMe Matter Institute EMMI, GSI Helmholtzzentrum f\"ur Schwerionenforschung GmbH, Darmstadt, Germany\\
$^{98}$ Saga University, Saga, Japan\\
$^{99}$ Saha Institute of Nuclear Physics, Homi Bhabha National Institute, Kolkata, India\\
$^{100}$ School of Physics and Astronomy, University of Birmingham, Birmingham, United Kingdom\\
$^{101}$ Secci\'{o}n F\'{\i}sica, Departamento de Ciencias, Pontificia Universidad Cat\'{o}lica del Per\'{u}, Lima, Peru\\
$^{102}$ Stefan Meyer Institut f\"{u}r Subatomare Physik (SMI), Vienna, Austria\\
$^{103}$ SUBATECH, IMT Atlantique, Nantes Universit\'{e}, CNRS-IN2P3, Nantes, France\\
$^{104}$ Sungkyunkwan University, Suwon City, Republic of Korea\\
$^{105}$ Suranaree University of Technology, Nakhon Ratchasima, Thailand\\
$^{106}$ Technical University of Ko\v{s}ice, Ko\v{s}ice, Slovak Republic\\
$^{107}$ The Henryk Niewodniczanski Institute of Nuclear Physics, Polish Academy of Sciences, Cracow, Poland\\
$^{108}$ The University of Texas at Austin, Austin, Texas, United States\\
$^{109}$ Universidad Aut\'{o}noma de Sinaloa, Culiac\'{a}n, Mexico\\
$^{110}$ Universidade de S\~{a}o Paulo (USP), S\~{a}o Paulo, Brazil\\
$^{111}$ Universidade Estadual de Campinas (UNICAMP), Campinas, Brazil\\
$^{112}$ Universidade Federal do ABC, Santo Andre, Brazil\\
$^{113}$ University of Cape Town, Cape Town, South Africa\\
$^{114}$ University of Houston, Houston, Texas, United States\\
$^{115}$ University of Jyv\"{a}skyl\"{a}, Jyv\"{a}skyl\"{a}, Finland\\
$^{116}$ University of Kansas, Lawrence, Kansas, United States\\
$^{117}$ University of Liverpool, Liverpool, United Kingdom\\
$^{118}$ University of Science and Technology of China, Hefei, China\\
$^{119}$ University of South-Eastern Norway, Kongsberg, Norway\\
$^{120}$ University of Tennessee, Knoxville, Tennessee, United States\\
$^{121}$ University of the Witwatersrand, Johannesburg, South Africa\\
$^{122}$ University of Tokyo, Tokyo, Japan\\
$^{123}$ University of Tsukuba, Tsukuba, Japan\\
$^{124}$ University Politehnica of Bucharest, Bucharest, Romania\\
$^{125}$ Universit\'{e} Clermont Auvergne, CNRS/IN2P3, LPC, Clermont-Ferrand, France\\
$^{126}$ Universit\'{e} de Lyon, CNRS/IN2P3, Institut de Physique des 2 Infinis de Lyon, Lyon, France\\
$^{127}$ Universit\'{e} de Strasbourg, CNRS, IPHC UMR 7178, F-67000 Strasbourg, France, Strasbourg, France\\
$^{128}$ Universit\'{e} Paris-Saclay, Centre d'Etudes de Saclay (CEA), IRFU, D\'{e}partment de Physique Nucl\'{e}aire (DPhN), Saclay, France\\
$^{129}$ Universit\'{e}  Paris-Saclay, CNRS/IN2P3, IJCLab, Orsay, France\\
$^{130}$ Universit\`{a} degli Studi di Foggia, Foggia, Italy\\
$^{131}$ Universit\`{a} del Piemonte Orientale, Vercelli, Italy\\
$^{132}$ Universit\`{a} di Brescia, Brescia, Italy\\
$^{133}$ Variable Energy Cyclotron Centre, Homi Bhabha National Institute, Kolkata, India\\
$^{134}$ Warsaw University of Technology, Warsaw, Poland\\
$^{135}$ Wayne State University, Detroit, Michigan, United States\\
$^{136}$ Westf\"{a}lische Wilhelms-Universit\"{a}t M\"{u}nster, Institut f\"{u}r Kernphysik, M\"{u}nster, Germany\\
$^{137}$ Wigner Research Centre for Physics, Budapest, Hungary\\
$^{138}$ Yale University, New Haven, Connecticut, United States\\
$^{139}$ Yonsei University, Seoul, Republic of Korea\\
$^{140}$  Zentrum  f\"{u}r Technologie und Transfer (ZTT), Worms, Germany\\
$^{141}$ Affiliated with an institute covered by a cooperation agreement with CERN\\
$^{142}$ Affiliated with an international laboratory covered by a cooperation agreement with CERN.\\

\end{flushleft} 

\end{document}